\documentstyle[12pt,a4,epsf,fleqn]{report}

\def\be{\begin{equation}}
\def\ee{\end{equation}}
\def\eq{\end{equation}}

\newcommand{\psbild}[5]
{\par
 \begin{figure}[#1]
 \begin{minipage}{0cm} \end{minipage}
 \hfill
 \begin{minipage}{#3}
 \refstepcounter{figure}\label{#2}
 \epsfxsize=#3
 \epsffile{#4}
 \end{minipage} 
 \hfill
 \begin{minipage}{0cm} \end{minipage}
 \parbox{14cm}{\centerline{\small {\bf fig. \ref{#2}:} #5}}
 \end{figure} }

\author{S\o{}ren K\"o{}ster \\  DTP 94-33 \\ hep-th/9408140}
\title{``Two dimensional QCD is a string theory'' \newline
Review on references \cite{DG} and \cite{GT1} }

\begin{document}
\maketitle
\begin{abstract}
This text, written as dissertation within the M.Sc. course in particle
theory
at the Centre for Particle Theory, University of Durham, during the
academic year 1993/94, reviews two articles by D.Gross and by D.Gross
and W.Taylor which interpret the 1/N- expansion of the partition
function  of $QCD_2$ as string perturbation series.
For this required mathematical and physical background is presented.

\end{abstract}



\section*{Foreword}
 
The task set for this dissertation was to give a review on the recent
development of an interpretation of pure Quantum Chromodynamics in two
dimensions ( $QCD_2$ ) as a string theory.
 
\vspace{1em}
 
The work in this direction was started by D.J. Gross \cite{DG}
, inspired, presumably, by an article of G. 't Hooft \cite{tH1}
, and is still in progress.
 
\vspace{1em}
 
The mathematical and physical concepts used for performing the
underlying idea that indeed $QCD_2$ can be discribed as a string
theory
stem from group theory, the theory of Riemann surfaces, loops in
algebraic
topology and lattice gauge theories, and, of course, from Yang-Mills
and string theories.
 
\vspace{1em}
 
A complete treatment of the progress done up to this time proved to be
impossible because of the extensive literature published very
recently and using tools I could not be familiar with at the start of
my work.
 
\vspace{1em}
 
Rather than to give an overview of all achievements I decided to
introduce as thoroughly as possible the concepts, tools and ideas used
in the main references. I will, therefore, follow the development
through
the article of D.Gross \cite{DG}, the first article by D.Gross and
W.Taylor \cite{GT1}, and the article by J.Minahan \cite{JM}. As
motivation for the relevance of planar diagrams and two dimensional
pure gauge theory for the discription of the strong interaction and
its
main characteristic, the confinement, I will give G.'t Hooft's
1/N-expansion \cite{tH1} and his derivation of a ``meson'' spectrum
\cite{tH2} .
 
\vspace{1em}
 
Additionaly the dissertation contains short accounts of the necessary
results
{}from group theory, algebraic topology, Riemann surfaces in particular,
and  lattice gauge theory.

\tableofcontents
 
\chapter{}
 
\section{Introduction}
 
The purpose of this chapter is to give reasons for seaking another
perturbation theory/ quantisation procedure for the strong interaction
( section 1.1 ), to connect the resulting 1/N-approach to planar
diagrams
( $QCD_2$, section 1.2 ), and to give an example why the physics
contained in this concept might be nontrivial and physically true: 't
Hooft's
derivation of a ``meson''spectrum ( section 1.3 ).
 
\vspace{1em}
 
By now the perturbation theory of QCD as a weak coupling expansion
seems to discribe correctly those contributions to high-energy
scattering, which
are due to the strong interaction.
 
Although this QCD is very sucessful in the high-energy- (UV-) limit,
it
fails to discribe the gauge theory in the infrared- (IR-) limit: The
renormalised coupling stops being small at ``large'' distances (ca.
1fm).
This is qualitatively in good agreement with the experimental
``observation''
that no single quarks or coloured states appear in the detectors, but
it does not provide a quantitative description of QCD in the IR, e.g.
to caculate masses and exitations of hadrons and thereby to check QCD
at the other end of momentumscale.
 
This is in spite of the fact that the quark model was invented to
analyse the huge spectrum of boundstates  observed as resonances in
high-energy experiments and that non abelian gauge theories were
proposed
to describe confinement.
 
\vspace{1em}
  
The spectrum of mesons is devided into families with linear
correlation
of squared mass and spin ( Regge-trajectories, see e.g. \cite{EJS} ).
Of course we would like to have a theory to calculate this spectrum,
even more because it appears to be simpler than the atomic spactra
which lead to the development of Quamtum Mechanics. More generally the
situation is very much the same as in the beginning of this century:
Without an analytic model for the spectrum the desciption of nature is
unsatisfactorily incomplete.
 
\vspace{1em}
 
Lattice gauge theory
has already provided contributions to the understanding of QCD in the
IR
(see eg. \cite{Wils} \cite{Ro} \cite{Creu}). There was made the
observation that in the strong coupling expansion (cf. section 2.1)
the free energy can be expressed as a sum over surfaces with quite
complicated contact terms, though (see references in \cite{DG} under
``[2]''). D. Gross takes this as an ``existence proof'' of a string
formulation of QCD \cite{DG} . But lattice gauge theories can not
provide an analytic description of QCD.
 
\vspace{1em}

In contrast to the gauge coupling, which gives no good expansion due
to
its growth at low momentum, there is hope, that there exists a
perturbation theory with $( \pi N )^{-1}$ as effective expansion
parameter
\cite{Mig1}. This corresponds to regarding $N = 3$ as ``large''; the
``limit'' $N \rightarrow \infty$ which is frequently used to determine
the right $1/N$ expansion of the partition function has to be
understood in this sense.
 
\vspace{1em}
 
 In the recent decades ``string theories'' have been proposed as
theories of the strong interaction (``dual models'') and as ``theories
of everything''. Although a lot is known about those theories it seems
that there is not much, by now, whose description is actually given by
a
string theory \cite{Green} .
 
Planar diagrams as in 't Hooft's paper \cite{tH1} and strings share
the property of being Riemann surfaces (or triangulations of those,
respectively) \cite{D'H} .
 
Both the representation theory of the gauge groups $SU(N)$ and $U(N)$
and the classification of Riemann surfaces can be connected to the
representation theory of the symmetric group of permutations $S_n$
(see chapters 3 and 5). The partition function of $QCD_2$ is solved
exactly as ``Fourier-''series over representations ( see chapter 2,
\cite{Mig2}, \cite{Rus} ).

\vspace{1em}
 
To summarise there is a necessity to find a theory for the IR-limit ,
there is hope that a planar expansion with $1/N$ as perturbative
parameter might be the concept to be applied, there are tools to
connect $QCD_2$ with a ``string'' perturbation theory, and there is
the hope that by the knowledge about string theories we can find a
string field theory giving this expansion and which leads to some
theory in higher dimension.
 
\vspace{1em}
 
 
In this context I will try to present the work of D.Gross and D.Gross
\& W.Taylor published in \cite{DG}\cite{GT1} alongside mathematical
and physical background.

Still within this chapter the particular limit for the 1/N- expansion
chosen by D.Gross is motivated following 't Hooft's work
and argumentation is given that two dimensional gauge models have
physically relevant features like confinement and a linear spectrum:
't Hooft's 2D meson model is used to argue that considering two
dimensions could teach lessons about higher dimensions.

Chapter 2 develops the exact solution of $QCD_2$ using the heat kernel
action.

Chapter 3 gives the necessary concepts and results from group theory
used in the later chapters.

Some discussion on one of the central terms in this approach, the
``Wilson loop average'', is given in chapter4.

Chapter 5 presents the tools from the theory of Riemann surfaces
required lateron.

The results achieved then are used to give the first step of the
interpretation: The understanding of the 1/N- expansion of the
$QCD_2$- partition function as a ``chiral'' sum over covering maps
{}from strings onto the $QCD_2$- spacetime manifold. (chapter 6)

The 1/N- expansion is improved for the limit $N \rightarrow \infty $
by introducing two coupled ``chiral'' sectors in chapter 7. The ``non
chiral sum'' is interpreted then.

Chapter 8, finally, outlines the treatment of bordered surfaces
 in this framework.

\section{'t Hooft's $U(N)$ model}
 
The purpose of this section is to present a gauge model showing
domination of planar diagrams in the limit  $N \rightarrow \infty $
and indicating
which specific $N \rightarrow \infty $ limit one has to take to find
the ``right'' $1/N$ expansion for eg. $QCD_2$. In the next section a
two dimensional version of this model will be considered to derive an
approximately linear spectrum of some ``$\mu^2$''.
 
\vspace{1em}
 
Because the usual generators of the Lie algebra {\sc u}(N) of U(N)
do not have  to be traceless as those of  {\sc su}(N) the Feynman
rules give rise to a simpler group theory  (``colour'') factor for
``index''
(coloured) loops. In fact it is simply N , which simplifies the
treatment of the model considerably.
 
The parameter $N$ is treated to be free ($N \rightarrow \infty $)   in
order to determine a reasonable 1/N expansion under the assumption
that $N=3$ is a large number.
 
In the following first step the quantisation of the $U(N)$ model is
outlined and Feynman rules are given to the required extend. The
second next step is to relate higher order diagrams to their
dominantly
planar character in the $N \rightarrow \infty $ limit, and to give the
specific limit used by D.Gross \cite{DG}.

\subsection{Feynman rules}
 
The gauge model chosen is U(N) gauge theory. The similarities and
differences between this model and QCD ( SU(N) model) can be extracted
{}from
a comparison of the Lie algebras  {\sc u}(N) and  {\sc su}(N).
 
{\sc u}(N) and {\sc su}(N) both consist of anti hermitean matrices and
satisfy the same algebra:
\cite{Oku}\cite{Corn}
\begin{equation}
[ A_i^j , A_k^l] = \delta_k^j A_i^l - \delta_i^l A_k^j
\end{equation}
 
where {\sc su}(N) consists of traceless matrices only i.e.
$
\mbox{{\sc u}(N)} \cong \mbox{{\sc u}(1)} \oplus \mbox{{\sc su}(N)}
$
and we have one abelian gauge field more for {\sc u}(N).
 
Lie algebra valued fields, i.e. gauge fields and ghosts, carry in both
models
two colour ``currents''; these are the same as the two colour lines of
gluons in QCD. With the standard choice of a basis for  {\sc
u}(N) \cite{Corn} \cite{Oku} each index stands for one out of N
``currents''.

The  perturbation theory and the Feynman rules follow, formally, after
the temporary introduction of external sources associated with each
field
(see \cite{Ram}) from the equation
\begin{equation}
Z( \{J_\omega\} ) =
e^{-S_I(-\frac{\delta}{\delta J_A},-\frac{\delta}{\delta
J_{\bar \psi}},-\frac{\delta}{\delta J_\psi
},-\frac{\delta}{\delta J_{\bar \eta}},-\frac{\delta}{\delta
J_\eta})} Z_0(J_A,J_{\bar \psi},J_\psi,J_{\bar \eta},J_\eta)
\end{equation}
 
This equation holds in euclidean spacetime. $Z_0$ is the partition
function given by the path integral weighted by the quadratic terms of
the effective Lagrangean only (\cite{Bai} sect.10.5). This path
integral can be solved in analogy with the following Gaussian integrals in
finite dimensions:
\newpage
\begin{displaymath}
\int d{\bar x}_1dx_1d{\bar x}_2..dx_n exp(-x^+\mbox{\bf A}x + \eta^+x
+ x^+\eta)
\end{displaymath}
\begin{equation}
 = - det(\mbox{\bf A}) exp(\eta^+\mbox{\bf A}^{-1}\eta)
\end{equation}
for anticommuting variables
\begin{displaymath}
\int d{\bar z}_1dz_1d{\bar z}_2..dz_n exp(-z^+\mbox{\bf B}z + \rho^+z
+ z^+\rho) 
\end{displaymath}
\begin{equation}
=(2 \pi)^n \ |det(\mbox{\bf B})|^{-1} exp(\rho^+\mbox{\bf B}^{-1}\rho)
\end{equation}
for commuting variables
 
$S_I$ contains the remaining terms which in turn are drawn outside the
the path integral like parameter differentiations of the
partition function
in statistical mechanics.

The external sources share the functional features of the fields they
are associated with, but they are external. For the moment this spoils
the gauge invariance, but the objects to be calculated will be gauge
invariant after the limit $J_\omega \rightarrow 0$, so the formalism
will be consistent after this limit \cite{Tay}.
 
After the choice of a unique and complete set of indices for each
family of functions, eg a basis for the Lie algebra, the conservation
laws
at the vertices follow formally from \cite{Gross}
 
\begin{equation}
\frac{\delta J_\omega ( x^{'} )}{\delta J_\Omega (x)} = \delta_{\omega
, \Omega} \
\delta ( x^{'} - x )
\end{equation}
 
So with the usual choice for   {\sc u}(N)  \cite{Corn},\cite{Oku} the
propagators of the U(N) model carry for the gauge field and the ghosts
two currents of one out of N colours each. These currents pass
vertices independently. By this any current/index loop produces a
colour factor
$\sum_i \delta_i^i = N$ for the amplitude of the diagram. This is
simpler than for QCD and the reason for using U(N) rather than SU(N).

  The possibility of assigning two independent currents to the gauge
and
ghost fields makes planar diagrams as considered in section 1.2.2
orientable.
 
\vspace{1em}
 
The additional abelian gauge field associated with   {\sc u}(1) in
{\sc u}(N)   does not affect the nonabelian character of the algebra,
so the model should give the right implications in the confining
IR-limit.
 
\vspace{1em}
 
't Hooft formulates the model in euclidean spacetime, i.e. after a
``Wick-''rotation, which does not affect the validity of the theory
nor changes the rules (or their derivation) significantly.
't Hooft chooses the Feynman gauge, such that the rules look like the
usual ones from QCD (see \cite{Bai} \cite{Ryd}).
\vspace{1em}
 
We do not need the Feynman rules in detail (see \cite{tH1}). Their
geometric character and the power of the coupling constant $\kappa$
attached to the vertices will be sufficient for the dicussion in
section 1.2.2.

\psbild{h}{figure1}{5cm}{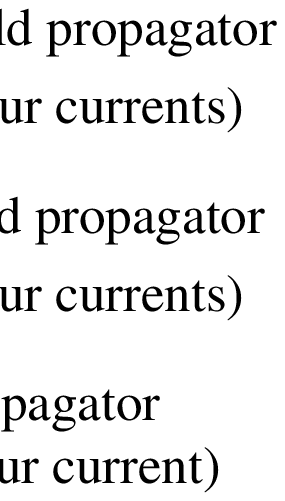}{the propagators}

\psbild{h}{figure2}{8cm}{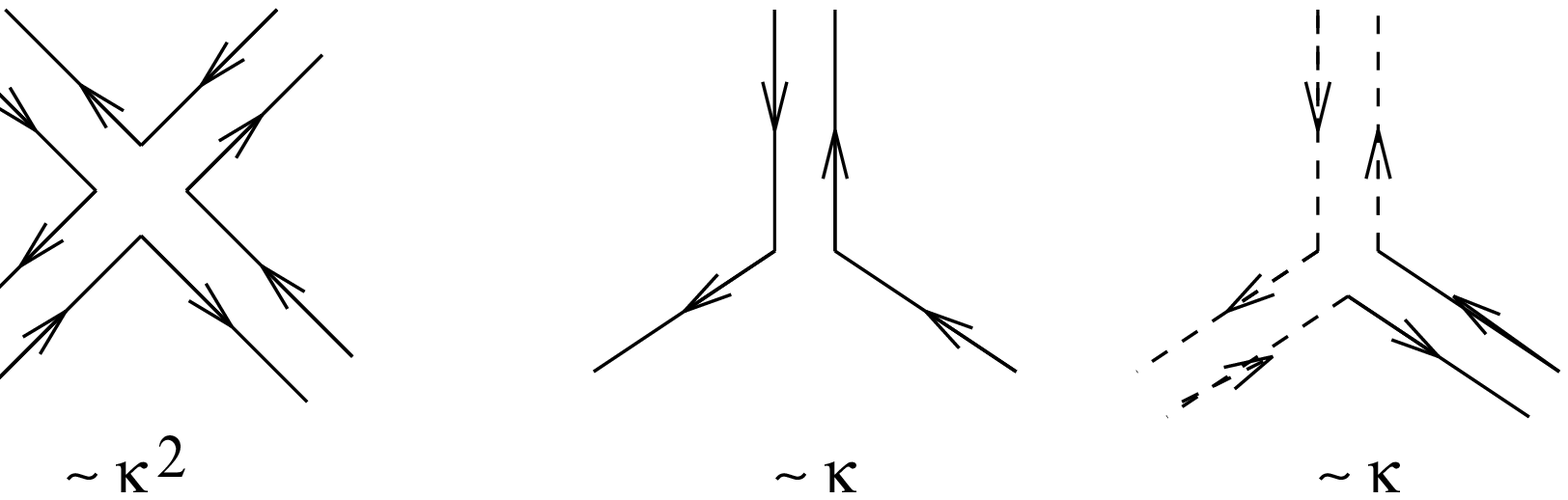}{the vertices}
 
\subsection{Planar diagrams \& large N limit}
 
Here we will consider connected diagrams of higher order in the gauge
coupling
$\kappa$
corresponding to the IR-limit, where $\kappa$ is no longer small
enough to describe the theory correctly with diagrams involving a few
vertices only.
 
We want to treat the diagrams as geometrical entities under the
assumption that the powers of N i.e. the number of index loops
associated with
a diagram by the Feynman rules make an implication on their importance
in the IR-limit. Because we are still dealing with the expansion in
powers of $\kappa$ we have to recognise the associated power of
$\kappa$,
too.
 
For this the external legs of the diagram are removed by gauge
invariant source functions, eg. ($f , f^{'} $ are flavour indices)
\begin{displaymath}
J \ = \ \sum_i {\bar \psi}^{f i} \  \psi^{f^{'}}_i
\end{displaymath}
By this there is no colour entering or leaving the diagram; the colour
flow
continues at $J$. The source looks like the creation/annihilation of a
meson (see figure \ref{figure3}).

\psbild{h}{figure3}{2cm}{}{source function}
 
Now we take the diagram and interpret each propagator as an edge,
gauge and ghost field propagators are internal edges: Every index-loop
is a face. The resulting polyhedron is finished by taking each
quark-loop
as a face, too.
 
For the planar diagrams  this gives a surface, for which the following
calculation holds:
 
Be F, E, V, L, I the numbers of faces, edges, vertices, quark loops,
and index loops in the diagram. Be $V_n$ the number of n-point
vertices.
 
Then are, by construction:
\begin{displaymath}
 F = L + I  \qquad   V = \sum_i V_i \qquad 2 E \ = \ \sum_n \ n V_n
\end{displaymath}
 
Each index loop contributes a factor N, each 3-point vertex a factor
$\kappa$, each 4-point vertex a factor $\kappa^2$.
 
This amounts to an overall factor associated to the diagram reading
as:
\begin{displaymath}
r \ = \ \kappa^{V_3 + 2 V_4} N^I \ = \ \kappa^{2 E - 2 V} \ N^{F - L}
\end{displaymath}
 
Because the surfaces are orientable the
Euler-Poincar\'e-characteristic $\chi$ is given by
 \begin{equation}
\chi \ = \ V - E + F \ = \ 2 - 2 g  \qquad \qquad
\mbox{\cite{Naka}:Theo.2.49}
\end{equation}
where $g$ is the number of holes (``genus'') in the surface of which
the diagram is the triangulation.
 
This leads to
\begin{equation}
r \ = \ ( \kappa^2 N)^{\frac{1}{2} V_3 + V_4} \ N^{2 - 2g - 2L}
\end{equation}
 
Under the fundamental assumption that a reasonable 1/N-expansion for a
gauge model like QCD exists this formula implies the following:
\begin{itemize}
\item Quark loops are comparatively suppressed for $N \rightarrow
\infty $
\item Planar diagrams arrange themselves according to their genus
\item The limit $N \rightarrow \infty$ and $\kappa^2 N = const.$
might lead to a good 1/N-expansion.
\end{itemize}
 
\psbild{h}{figure4}{7cm}{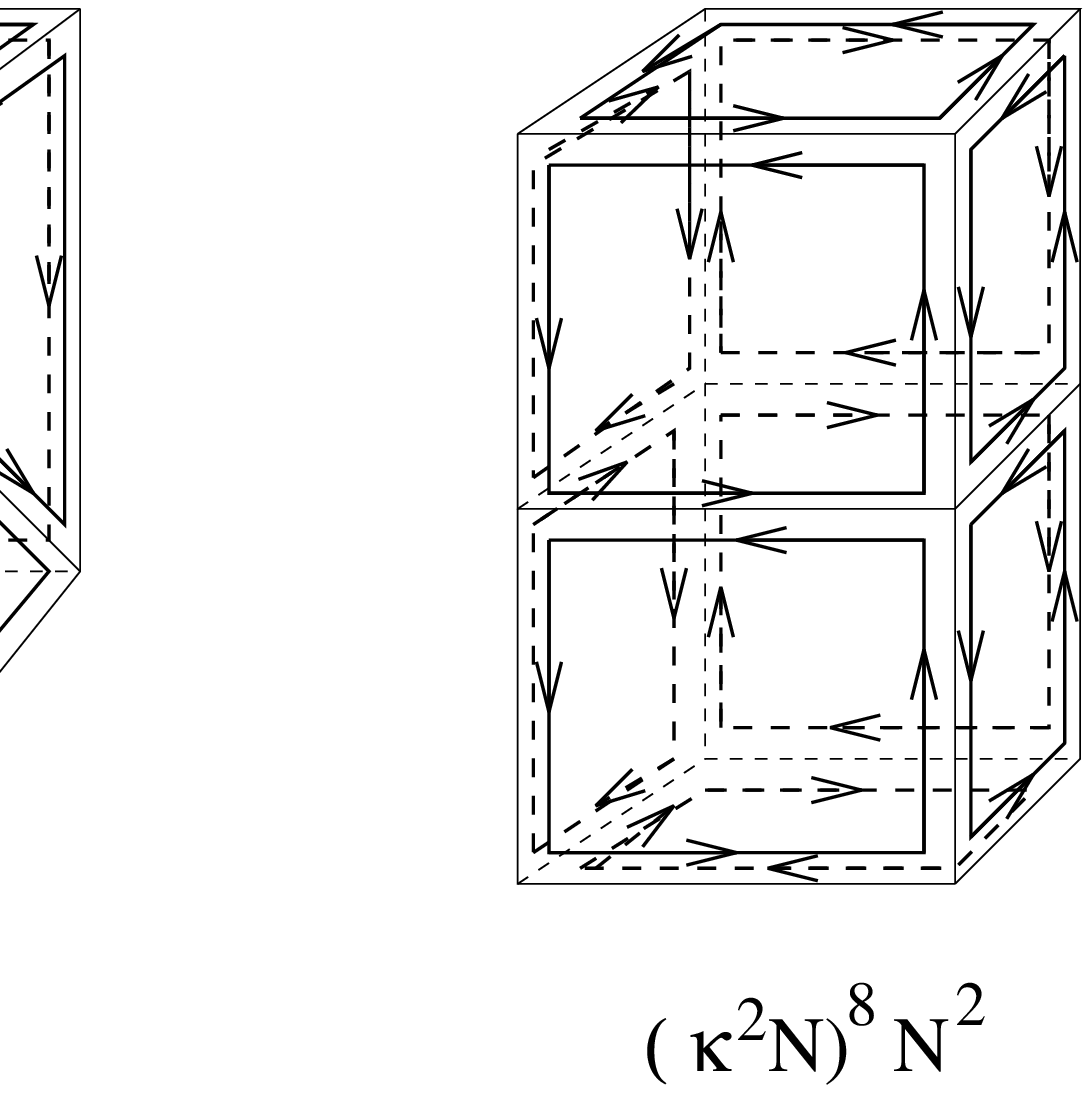}{comparison planar and nonplanar diagram}
 
\newpage

Regarding the power of N and $\kappa$ associated with a diagram as a
weight for its contribution leads to the conclusion, that planar
diagrams are dominating when $N \rightarrow \infty $  . It should be
sufficient to examine a simple, nontrivial example of two diagrams
with the same set of vertices in d=3, one is planar, one is not
(see figure \ref{figure4}). This shows that nonplanar diagrams are
suppressed because the flow
through the ``interior'' of the diagram admits less loops.

\vspace{1em}
 
The indication that considering planar gauge theories could be useful
will be supported  furtherly in the next section.

\section{Two dimensional $U(N)$ meson model}
 
In the following treatment of an ``amusingly simple model'' ('t
Hooft \cite{tH2}) two characteristics of strong interaction appear:
Single quarks have infinite mass (confinement) and and there is a
discrete, linear spectrum connected with quark pairs (Regge-
trajectories see eg.\cite{EJS})
 
The procedure followed in this section will be such:
\begin{itemize}
\item in 1.3.1 the very convenient Feynman rules are developed
\item section 1.3.2 solves the only dressed propagator 
\item and finally, in 1.3.3, we see approximate solutions of an
 hermitean eigenvalue equation, which one could regard as something
like the Klein-Gordon equation.
\end{itemize}

\subsection{Feynman rules}
 
The model is a U(N) gauge theory in two euclidean dimensions.
 
The Lagrangean leading to the Feynman rules is the usual:
\begin{displaymath}
\mbox{\it L} \ = \ \frac{1}{4} \ \mbox{\it F}_{\mu \nu  i}^j \mbox{\it
F}_{\mu \nu j}^i \ - \ {\bar \psi}^{fi}(\gamma_\mu \mbox{\it D}_\mu \
+ \ m_{(f)} )\psi^f_i
\end{displaymath}
\begin{displaymath}
 \mbox{\it F}_{\mu \nu i}^j \ = \ \partial_\mu A_{i \nu}^j -
\partial_\nu A_{i \mu}^j \ + \ \kappa [ A_\mu , A_\nu ]_i^j
\end{displaymath}
\begin{displaymath}
 \mbox{\it D}_\mu \psi^f_i \ = \ \partial_\mu \psi^f_i \ + \ \kappa
A_{i \mu}^j \psi^f_j
\end{displaymath}
\begin{displaymath}
A_{i \mu}^j (x) \ = \ -A_{j \mu}^{* \ i} (x)
\end{displaymath}

The indices i, j are Lie algebra indices, $\mu , \ \nu$ are spacetime
indices, f is the flavour index, $\psi$ quark fields, A the
antihermitean gauge field.
 
\vspace{1em}
 
There are two euclidean coordinates on the plane: $x_0, \ x_1$.
Lightcone coordinates $x_\pm \ = \
\frac{1}{\sqrt{2}} (x_1 \ \pm \ x_0) $ prove to be convenient.
 
A notion of upper and lower indices and a summation convention are
introduced  as follows:
\begin{displaymath}
x_1 \ \equiv \ x^1 \qquad x_0 \ \equiv -x^0
\end{displaymath}
\begin{displaymath}
x^\mu p_\mu  \equiv \ x^\mu p^\mu \equiv \ x_\mu p_\mu  \equiv \
x_+p^+ + x_- p^-  \equiv \ x^+ p^- + x^- p^+ \equiv \ x_+ p_- +
x_- p_+
\end{displaymath}

The algebra for the $\gamma$-matrices may be determined using the
lightcone versions of the Dirac- and
the Klein-Gordon- equation (as usual) and is:
 
\begin{equation}
\gamma_-^2 \ = \ \gamma_+^2 \ = \ 0 \qquad \{ \gamma_+ , \gamma_- \} \
= \ 2
\end{equation}
 
\vspace{1em}
 
The gauge condition is an axial gauge, the lightcone gauge:
$A_- \ = A^+ \ = 0$
 
Which leads to $ F_{+ -} = -\partial_- A_+ $ and
 
\begin{displaymath}
\mbox{\it L} \ = \ -\frac{1}{2} tr (\partial_- A_+ )^2 \ - {\bar
\psi}^f (\gamma \partial \ +m_{(f)} \ + \kappa \gamma_- A_+) \psi^f
\end{displaymath}
There is no coupling to ghosts in this gauge (cf. \cite{Ryd}).
 

\newpage
 
\psbild{h}{figure5}{8cm}{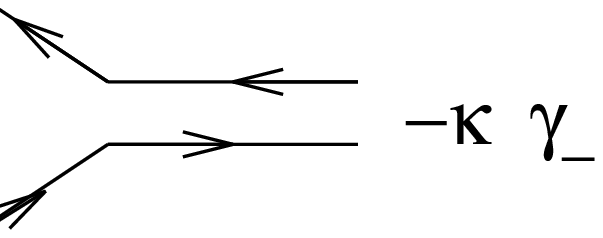}{Feynman rules for meson-model}


Because of the simple $\gamma$-algebra and the fact that there is only
a single vertex in the theory the $\gamma$-matrices disappear from the
theory whenever vertices are connected to propagators. A simple
calculation leads to an equivalent set of simpler rules (see figure 
\ref{figure6}).


\psbild{h}{figure6}{8cm}{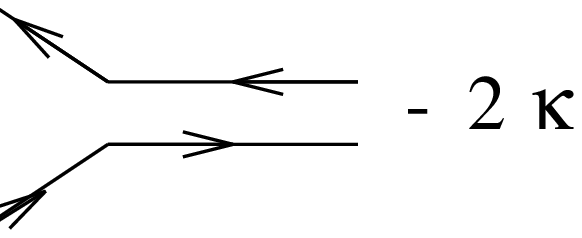}{effective Feynman rules}
 
In the following we will ignore contributions from quark loops in
agreement with the observation
in the previous section.

\newpage
\subsection{{\sc The \/} dressed propagator}

\psbild{h}{figure7}{7cm}{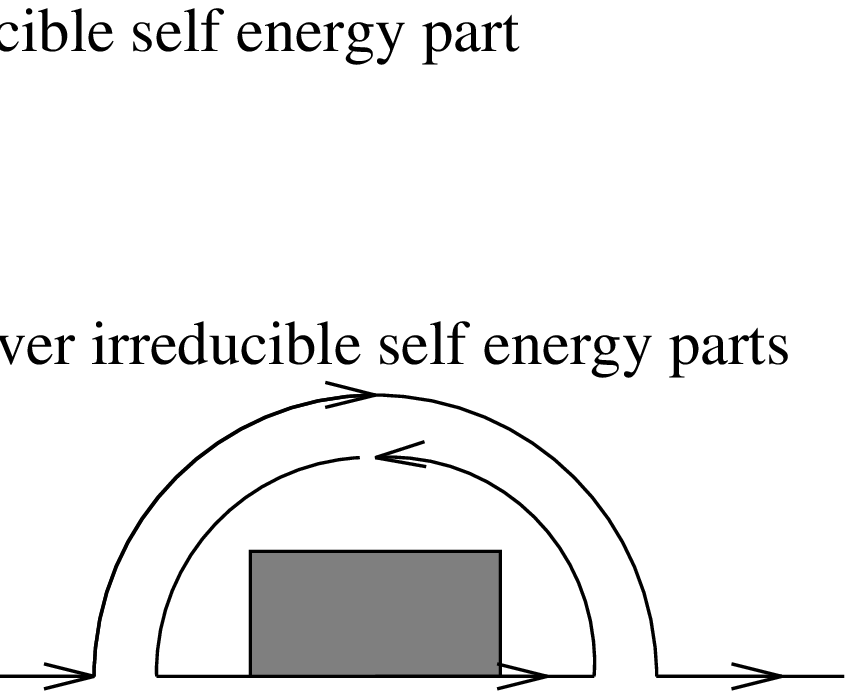}{equation for dressed propagator}
 
As usual we need the dressed propagators of the theory, i.e. two point
Green's functions including all self energy parts of the particle
(cf.\cite{Bai}, \cite{Ryd})
. The
gauge field propagator is already ``dressed'' because we have no
coupling between gauge fields and we neglect quark loops. So {\em the
\/} only
one to be dressed is the quark propagator.
 
{}From the form of the undressed propagator we already see the form of
the dressed propagator:
\begin{displaymath}
\frac{-i k_-}{m^2 \ +2k_+k_- \ -k_- \Gamma (k) \ -i\epsilon}
\end{displaymath}
where $i\Gamma (k)$ stands for the sum of one-point-irreducible self
energy parts.
 
The model is very simple and we can find  $i\Gamma (k)$ by solving a
quite simple equation it {\em has \/} to satisfy (see figure \ref{figure7}).

\begin{displaymath}
i\Gamma (p) = \frac{4 \kappa^2}{(2 \pi)^2 i}
\int dk_+ dk_- \frac{1}{k_-^2} \frac{-i (k_- +p_-)}{m^2 + [2(k_+ +
p_+)- \Gamma (k+p)] (k_- +p_-) -i\epsilon}
\end{displaymath}
 
We observe that $k_+$, $p_+$ only appear in the constellation $k_+ +
p_+$. The integral will be invariant under the shift  $k_+ +
p_+ \rightarrow k_+$ after, possibly, a regulation business, such that
$\Gamma (p)$ is independent of $p_+$.
 
The UV-divergence in the $k_+$ integral, as it can be seen easily by
choosing a symmetric cutoff, factorises. This factor is not 
involved with the following calculations and can be ignored henceforth
(cf.section 2.2).
 
The $k_+$-integral is, up to the divergence, given by:

\begin{displaymath}
\frac{\pi i}{2 |k_- \ +p_-|}
\end{displaymath}
This gives for the $k_-$-integration an IR-divergence. The choice of a
symmetric cutoff $\pm \lambda$ leads to a cancellation of all $\lambda$
{}from
all formulae at a later stage of the calculation; so presumably its
choice does not affect the final result.
\vspace{1em}
 
Performing the  $k_-$-integral in this manner leads to:
\begin{equation}
\Gamma (p) = \ \Gamma (p_-) = \ -\frac{\kappa^2}{\pi} (\frac{\sigma
(p_-)}{\lambda} \ -\frac{1}{p_-})
\end{equation}
$\sigma(p_-)$ is the sign of $p_-$
 
 And the dressed propagator is:
\begin{equation}
\frac{-i k_-}{m^2 \ - \frac{\kappa^2}{\pi} \ + 2  k_+ k_- \ 
+\frac{\kappa^2 |k_-|}{\pi \lambda} \ -i\epsilon}
\end{equation}
\vspace{1em}
 
In the limit $\lambda \rightarrow 0 $ the pole of the propagator is
shifted towards $k_+ \rightarrow \infty $. This means that there are
no free quarks in this model: They would have infinite mass.

As mentioned above the cutoff $\lambda$ will not be involved in the
``meson''spectrum calculated below. This means that quarks are
confined to the neighbourhood of (an-)other quark(-s).
 
\newpage
 
\subsection{Derivation of the approximate spectrum}

\psbild{h}{figure8}{5cm}{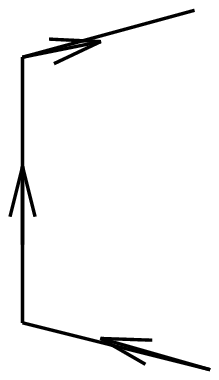}{equation for meson spectrum}
 
For ``blobs'' out of which a quark with mass $m_1$ and momentum $p$
and
an antiquark with mass $m_2$ and momentum $r-p$ appear exists, due to
the
simple
Feynman rules of this model, an equation, depicted in figure
\ref{figure8}, 
which will
lead to the spectrum. $\psi$ stands for such an arbitrary vertex about
which we do not need to know much more than that is has to satisfy the
equation below.

\begin{displaymath}
\psi (p,r) = \ -\frac{4\kappa^2}{(2\pi)^2 i} \quad \times
\end{displaymath}
\begin{displaymath}
(p_- - r_-)[M_2^2 + 2(p_+-r_+)(p_- -r_-) + \frac{\kappa^2}{\pi
\lambda} |p_- - r_-| -i\epsilon]^{-1} \quad \times
\end{displaymath}
\begin{displaymath}
p_- [M_1^2 + 2 p_+p_- + \frac{\kappa^2}{\pi
\lambda} |p_-| -i\epsilon]^{-1} \quad \times
\end{displaymath}
\begin{equation}
\int\!\!\!\int dk_+ dk_- \frac{\psi (p + k, r)}{k_-^2}
\end{equation}
\begin{displaymath}
M_i^2 = m_i^2 \ - \frac{\kappa^2}{\pi}
\end{displaymath}
 
\newpage

Writing $\phi (p_-, r) = \int dp_+ \psi (p_+, p_-,r)$ leads to:
\begin{displaymath}
\phi (p_-,r) = -\frac{\kappa^2}{(2\pi)^2 i} \quad \times
\end{displaymath}
\begin{displaymath}
\int dp_+ [p_+ - r_+ +\frac{M_2^2}{2(p_- -r_-)} +
(\frac{\kappa^2}{2\pi \lambda} -i\epsilon) \ \sigma(p_- - r_-)]^{-1} 
\quad \times
\end{displaymath}
\begin{displaymath}
[p_+ + \frac{M_1^2}{2p_-} + (\frac{\kappa^2}{2\pi \lambda} -i\epsilon)
\ \sigma(p_-)]^{-1} \quad \times
\end{displaymath}
\begin{equation}
\int dk_- \frac{\phi (p_- +k_-, r)}{k_-^2}
\end{equation}

The integral can be evaluated by summing over the poles in the upper
half plane according to:
\begin{equation}
\int \limits_{-\infty}^{+\infty} dx R(x) \ = \ 2 \pi i \sum_{Im(z)>0}
res_z R \hfill \mbox{\cite{FiLi}p.152,S.6.1} \qquad
\end{equation}
 
In case {\em both \/} poles lie in the upper half plane, the
sum is always zero: A simple calculation shows, that the two residua
cancel each other. So the nontrivial case is when the poles lie on
different sides of the real axis, i.e. $ \sigma(p_-) = \ -\sigma(p_- -
r_-)$.
We may take $r_- > 0$.
 
Then:
\begin{displaymath}
\phi (p_-,r) = \frac{\kappa^2}{2\pi} \ \Theta (p_-) \Theta (r_- -
p_-) \quad \times
\end{displaymath}
\begin{displaymath}
[\frac{M_1^2}{2p_-} + \frac{M_2^2}{2(p_- -r_-)} + \frac{\kappa^2}{\pi
\lambda} + r_+]^{-1} \quad \times
\end{displaymath}
\begin{equation} \label{Thetas}
\int dk_- \frac{\phi (p_- +k_-, r)}{k_-^2}
\end{equation}
where the usual theta-functions $\Theta (x)$ keep track of the position
of the poles.
 
\newpage

Using the principal value $\wp$ and the Sokhotsky-Plemelj-formula
\cite{Flue} \cite{Florin}:
\begin{equation}
\wp \int dk_- \frac{\phi (k_-)}{k_-^2} \ = \ \frac{1}{2} \int dk_-
\frac{\phi (k_- + i\epsilon)}{(k_- + i\epsilon)^2} +  \frac{1}{2} \int
dk_-
\frac{\phi (k_- - i\epsilon)}{(k_- - i\epsilon)^2}
\end{equation}
\begin{equation}
\int dk_- \frac{\phi (p_- +k_-, r)}{k_-^2} \ = \ \frac{2}{\lambda}
\phi(p_-) + \wp \int dk_- \frac{\phi (p_- +k_-, r)}{k_-^2}
\end{equation}
 
Now we use all this and dimensionless variables:
\begin{displaymath}
\alpha_{1,2} = \frac{\pi M_{1,2}^2}{\kappa^2} = \frac{\pi
m_{1,2}^2}{\kappa^2} -1
\end{displaymath}
\begin{displaymath}
-2 r_+ r_- =  \frac{\kappa^2}{\pi} \mu^2 \qquad \frac{p_-}{r_-} = x
\end{displaymath}
which leads to
\begin{equation}
 \mu^2 \phi(x) \ = \ (\frac{\alpha_1}{x} + \frac{\alpha_2}{1-x}) \
\phi(x) - \wp \int\limits_0^1 dy \frac{\phi (y)}{(y-x)^2} \ \equiv
\mbox{\bf H} \phi
\end{equation}
 
Two observations have to be made:
\begin{itemize}
\item The dependence on the cutoff $\lambda$ does no longer {\em
appear \/} to exist.
\item The solutions of this equation should satisfy the equation
(\ref{Thetas})
it was derived from. We have therefore to examine the solutions which
vanish at 0 and 1 and have support only within [0,1].
\end{itemize}
 
Interpreting this equation as an eigenvalue equation of the operator
{\bf H} defined by this equation one can easily show that for two
solutions
$\psi, \ \phi$ we have:
\begin{displaymath}
(\psi, \mbox{\bf H} \phi) \ = \ (\phi,  \mbox{\bf H} \psi)
\end{displaymath}
(by application of the definition, the usual scalarproduct and the
observation about the support of $\psi, \phi$)
 
\newpage

Now we solve the equation approximately:
\begin{displaymath}
\wp \int\limits_0^1 dy \frac{e^{i\omega y}}{(y-x)^2} \approx \ \wp
\int\limits_{-\infty}^{+\infty} dy \frac{e^{i\omega y}}{(y-x)^2}
\qquad (x \in [0,1])
\end{displaymath}
\begin{displaymath}
\approx \pi i e^{i\omega x} \omega
\end{displaymath}
by using the definition of $\wp$ and the formula:
\begin{displaymath}
\int dx \ R(x) e^{ix} = \ 2\pi i \sum_{Imz>0} res_z (R(\zeta)
e^{i\zeta})
\qquad \mbox{\cite{FiLi}p153,S.6.2}
\end{displaymath}
 
For $\alpha_{1,2} \approx 0$ the following functions approximately
solve
the equation:
\begin{displaymath}
\phi^k (x) \simeq \sin k\pi x \qquad k \in I\!\!N -\{ 0 \}
\end{displaymath}
\begin{displaymath}
\Rightarrow \mu_k^2 \approx \pi^2 k
\end{displaymath}
 
So within the limits of the approximations made we have a linear,
discrete spectrum of an hermitean operator {\bf H}.
 
This model may serve as a motivation that we might carry nontrivial
physics through a treatment based on two dimensional
Yang-Mills-theory.

\chapter{The Heatkernel Action}
 
This chapter is devoted to the derivation of the exact solution of
Yang-Mills theory in two dimensions, especially for the case of
$QCD_2$.
 
The solution stems from concepts introduced for the lattice
approximation
of the theory. Therefore some backround from lattice gauge theory will
be given first.
 
\section{Wilson's action}
 
In the usual continuum Yang-Mills theory a gauge field is introduced
to
define a connection which makes the derivative covariant under gauge
transformations. The gauge field then tells how to transport vectors
along paths in spacetime.
 
Equally well one could start with a set of parallel transporters which
act on the vector along the path and determine the result of parallel
transportation from an initial state of the vector.
 
{}From the gauge field point of view these parallel transporters are
given by ``pathordered exponentials'' (\cite{Naka}p.337, \cite{Goe}).
The Lie algebra contains the generators of infinitesimal gauge
transformations. A parallel transporter can therefore be regarded 
as a sucessive
application of Lie algebra operators along a path C in the limit of
zero
steplength. The order of those sucessive operations has to be defined
for nonabelian algebras. This defines the ``pathordered exponential'',
which actually has charateristics of an exponential map:
\begin{displaymath}
\mbox{\bf P} exp( \ \int\limits_C \ \mbox{\bf A}_\mu dx^\mu \ )
\end{displaymath}
 
\vspace{1em}
 
On a lattice the path integral quantisation involves integration over
all gauge transformations at all links between each two neighbouring
points on the lattice. It does not make much sense to use gauge fields
as
the variables of this procedure. One rather uses the variables which
are
the actual variables of integration. For this an appropiate action has
to be found to weight the configurations in terms of these variables.
 
I wish give an example, Wilson's action, and its continuum limit to
illustrate that actions like this one define an approximation which
might be equivalent to the more usual continuum theory in the limit of
small
lattice spacing a. Lattice gauge theories therefore provide models for
cases where the weak coupling perturbation theory does not work.
 
Wilson's action is, like the heat kernel action, formulated in terms
of group transformations assiociated with each elementary closed path
on the lattice, i.e. around the ``plaquettes'' which are for a cubic
lattice
simply rectangles with edges of length a.
 
It is \cite{Itz1} \cite{Wils}
\begin{displaymath}
S_j = \ \beta_j \sum_p \frac{1}{N} \frac{1}{2} tr( \mbox{\bf U}_p +
\mbox{\bf U}_p^+) = \ \beta_j \sum_p \frac{1}{N} Re(tr( \mbox{\bf
U}_p))
\end{displaymath}
where N is the gauge group parameter, the sum is over all plaquettes
in the lattice and $\mbox{\bf U}_p$ is the paralleltransporter once
along the
edges
of the plaquette, sometimes called the ``holonomy'' (cf. section on
Riemann
surfaces).
 
$\beta_j $ is a free parameter which is chosen to give the right
continuum limit; regarding the action as a weight in the partition
function
of the lattice (like in solid state physics) $ \beta_j $ is called the
``temperature parameter''.
 
\vspace{1em}
 
The continuum limit is found by the observation, that the traces are
simultanuously maximal for the identity \cite{Itz1}. Cutting the
lattice into\
 a
maximal tree and the condition $\mbox{\bf U}_p = \mbox{\bf 1} \ \forall p$
leads to a configuration where all holonomies associated with an edge
are given by pure gauge.
 
$\mbox{\bf U}_{ij}$ be such an edge transformation, then there exist
two gauge transformations $g_i (g_j)$ at the points $i (j)$ on the
lattice such that:
\begin{displaymath}
 \mbox{\bf U}_{ij} = g_ig_j^{-1} \mbox{
\hfill \cite{Itz1}} \qquad
\end{displaymath}
\vspace{1em}
 
The further procedure is to perturb this configuration by
introduction of a gauge field, i.e.
\begin{displaymath}
\mbox{\bf U}_{ij} = g_i \mbox{\bf P} exp( \ \int_i^j \ \mbox{\bf
A}_\mu
dx^\mu \ ) g_j^{-1}
\end{displaymath}
and to use the Baker-Campbell-Hausdorff formula \cite{Wilc}
\cite{Mill} to derive $\mbox{\bf U}_p$. then the lattice spacing is
send
to zero, differences are replaced by derivatives, summations by
integrations leading to (in two dimensions)
\begin{displaymath}
S_j \approx \frac{\beta_j \kappa^2 a^2}{N} \int dx^2 (-\frac{1}{4
\kappa^2} tr(\mbox{\it F}_{\mu \nu}^{\ \ \ 2}))
\end{displaymath}
The factor in front of the integral has to be unity to give the right
continuum limit:
\begin{displaymath}
\beta_j = \frac{N}{\kappa^2 a^2}
\end{displaymath}
(For this reason the strong coupling expansion is ``natural'' for
lattice gauge theory \cite{Ro}.)

\section{Wilson loop average for $YM_2$}
 
Because the heat kernel action is determined in the continuum limit by
another object, the Wilson loop average, we first have to define this
object and to solve it in two dimensions.
 
The Wilson loop average is defined as the expectation value of
 
\begin{equation}
tr( \mbox{\bf P} exp( \ - \oint\limits_C dx^\mu \ \mbox{\bf A}_\mu
))
\end{equation}
This quantity is gauge invariant.
 
For the calculation below, following \cite{Paul}, we will use an axial
gauge, which can always be found by usage of the path ordered
exponential
(see eg. \cite{Itz2}p.566f.).
 
The Wilson loop average in two euclidean dimensions and free gauge
field is:
\begin{equation}
W[ C ] = \ \frac{1}{\mbox{\sc N}} \int [ \mbox{\it D \/}\mbox{\bf A }
] tr( \
  \
\mbox{\bf P} exp( \ - \oint\limits_C dx^\mu \ \mbox{\bf A}_\mu
))
\ exp \int \frac{1}{4 \kappa^2} tr(\mbox{\it F}_{\mu \nu}^{\ \ \ 2})
\end{equation}
 
Where {\sc N} stands for the usual normalisation, $ [ \mbox{\it D
\/}\mbox{\bf A } ] $ for the path integral and W[C] is the Wilson loop
average of the closed path C.
 
Taking a gauge transformation
\begin{displaymath}
g(x_1,x_2) = \ \ \mbox{\bf P} exp( \ - \int\limits_0^{x_2} dx^2 \
{\mbox{\bf A} (x_1,x_2)}_2
)
\end{displaymath}
which satisfies \cite{Naka}
\begin{displaymath}
g^{-1} \frac{\partial}{\partial x_2} g = \ -g^{-1} \mbox{\bf A}^2 g
\end{displaymath}
therefore we have
\begin{displaymath}
 \mbox{\bf A}^{' 2}= \ g^{-1}  \mbox{\bf A}^2 g \
+g^{-1}\frac{\partial}{\partial x_2} g = \ 0
\end{displaymath}
 
The field strength is then given by $ tr( \partial_2 \mbox{\bf
A}_1)^2$.
 
The integral is solved in analogy with the integral
\begin{displaymath}
\int d{\bar z}_1dz_1d{\bar z}_2..dz_n exp(-z^+\mbox{\bf B}z + \rho^+z
+ z^+\rho) 
\end{displaymath}
\begin{displaymath}
= (2 \pi)^n \ |det(\mbox{\bf B})|^{-1} exp(\rho^+\mbox{\bf
B}^{-1\
}\rho)
\end{displaymath}
This leads to
\begin{equation}
W[ C ] = \ tr \mbox{\bf P}exp( \ -\frac{\kappa^2}{2} \oint\limits_C
\oint\limits_C
dx^1 dy^1 G(x,y) \mbox{\bf T}^a_{x^1} \mbox{\bf T}^a_{y^1})
\end{equation}
where G satifies:
\begin{displaymath}
\partial_2^{\ 2} G(x,y) = \ \delta (x,y)
\end{displaymath}
for which in commuting variables the solution is \cite{Gross}
\begin{displaymath}
G(x,y) = \ \frac{1}{2} | x^2 - y^2 | \delta(x^1 - y^1)
\end{displaymath}
 
The Green's function is ``singular'': it gives ``infinite''
contribution for any pair $x,y$ with $y^1 - x^1 = 0$. In this sense
the divergence of the $k_+$-integral in section 1.3.2. can be
understood.
 
$\mbox{\bf T}^a_{x^1}$ is a basis vector in the Lie algebra, which
carries a location index due to the path ordering.
 
We now evaluate the exponential for a rectangle (see figure \ref{figure9}).
 
 

\psbild{h}{figure9}{5cm}{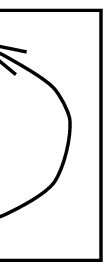}{evaluation of loop integral}
 
Along the $x_2$ edges we have no contribution and the path ordering
leads to $\mbox{\bf T}^a_{x^1} \mbox{\bf T}^a_{x^1} = C_2$ all
along the path, where $C_2$ is the quadratic Casimir (see next
chapter).
 
The integral is $(-1)$ times the area A of the enclosed rectangle.
 
This leads to:
\begin{equation}
W[ C ] = \ tr( exp( \ \frac{\kappa^2}{2} C_2 A \ ))
\end{equation}
 
So far there was no need to specify a representation R of the Lie
algebra.
If we choose a particular irreducible representation we have (see next
chapter):
\begin{displaymath}
W[ C ] =tr_R exp( \ \frac{\kappa^2}{2} C_2(R) A \ \mbox{\bf 1}_R)
  \\ = exp( \ \frac{\kappa^2}{2} C_2(R) A ) dim(R)
\end{displaymath}
$dim(R)$ is the dimension, $ C_2(R)$ the Casimir-eigenvalue of the
representation.
 
The next section will show that the choice of a rectangle implies no
restriction, because the $YM_2$-action is invariant under
area-preserving diffeomorphisms; so the ``area law'' stated above will
hold for all Wilson loops with no self intersections\footnote{Remark:
G.West \cite{West} found a relation between the Wilson loop average
and the gluon propagator $D_{\mu \nu}^{a b}$ in the general case:
\begin{displaymath}
W[C] \leq exp(  \frac{\kappa^2}{2} \oint\limits_C dx^\mu
\oint\limits_C dy^\nu D_{\mu \nu}^{a b} \delta_{a b} )
\end{displaymath}}.

\section{Additional symmetry of pure $YM_2$}
 
In two dimensions there is another symmetry of the free YM-action.
We follow the treatment given in \cite{Cor1}.
 
\begin{displaymath}
S_{YM} = \frac{1}{4 \kappa^2} \int\limits_{\sum} tr (\mbox{\bf F}
\wedge *\mbox{\bf F} )
\end{displaymath}
where $\sum$ is the spacetime surface, {\bf F} is the field strength,
{*\bf F} stands for the Hodge star of {\bf F} ,$\wedge$ is the wedge
product of differential forms (for definitions: \cite{Naka}).
 
Choosing an area form $\mu$ on $\sum$ one can write F with a Lie
algebra valued function $\phi$ as:
\begin{displaymath}
\mbox{\bf F}= \ \phi \mu
\end{displaymath}
and following the definition of the Hodge star
\begin{displaymath}
 \phi = *\mbox{\bf F}
\end{displaymath}
This is in components with $g$ being the determinant of the spacetime
metric:
\begin{displaymath}
F_{i j}^a = \ \sqrt{|g|} \epsilon_{i j} \phi^a
\end{displaymath}
$\epsilon_{i j}$ is the totally antisymmetric tensor (Levi-Cevita
symbol) in two dimensions, carries all of the tensorial character of
{\bf F} and thereby admits the invariance:
\begin{equation}
S_{YM} = \frac{1}{4 \kappa^2} \int\limits_{\sum} \mu tr(\phi \phi) =
S_{YM} = \frac{1}{4 \kappa^2} \int\limits_{\sum} dx^{\ 2} \sqrt{|det
g_{i j}|} tr\phi^2
\end{equation}
Because $\phi$ is a scalar any diffeomorphism $\sum \rightarrow \sum$
which leaves the area invariant will leave the action invariant.

\section{Solution of $QCD_2$}

\subsection{Orthogonality of characters}
 
Although some more group theory is yet to come we need some of it now.
 
The character of a group element is the trace of its representative in
some linear representation of the group. Of particular interrest are
characters in irreducible unitarian representations (cf. next
chapter).
 
Compact (or ``unimodular'') groups have the nice feature to admit an
invariant normalisable measure (the ``Haar''measure) such that
statements
{}from the representation theory of finite groups can often be
transfered to compact Lie groups using the correspondence
\begin{displaymath}
\frac{1}{|G|} \sum_{g \in G} \leftrightarrow \int dU \qquad  \qquad
\int dU = 1
\end{displaymath}
U(N) and SU(N) are unimodular(see \cite{Nach}).
 
The proof of the orthogonality and completeness of irreducible
characters given in \cite{Tung}(Th.3.5, 3.7) generalises and leads to:
\begin{equation} \label{orthochar}
 \int dU \chi_R (U) \chi_S (U^+) = \delta_{R S}
\end{equation}
 
Formulae proved in the same way are:
\begin{equation}
 \int dU \chi_R (V U) \chi_S (U^+ W) = \frac{\delta_{R S}}{dim(R)}
\chi_R (V W)
\end{equation}
\begin{equation}
 \int dU \chi_R (U V U^+ W) = \frac{\chi_R (V)\chi_R (W)}{dim(R)}
\end{equation}
 
The orthogonality  equation (\ref{orthochar}) is for totally
decomposable 
(into finite
dimensional irreducible unitarian) representations the completeness
statement that the
irreducible characters form a complete set on the space of continuous
class functions, which in turn lie dense in the set of all squared
summable classfunctions (see \cite{Serre} sect.4.3, \cite{GTM98}).
 
Class functions are functions of group elements which are invariant
under conjugation of the group element, i.e. depend only on the
conjugacy classes:
\begin{displaymath}
f( g ) \ = \ f( hgh^{-1}) \ = \ f( [g])
\end{displaymath}
 
By the Peter-Weyl theorem \cite{GTM129} each representation of compact
groups is totally decomposable into finite dimensional irreducible
unitarian representations (\cite{Serre} section4.3, \cite{GTM98}).
This means that there is a ``Fourier'' expansion for each class
function on U(N) and SU(N). By its gauge invariance every summable
function of the Wilson loop average can be expressed as a
``Fourier''series.
 
\vspace{1em}
 
This and the orthogonality of the characters are the two mathematical
building blocks to solve $QCD_2$  exactly. The physical contribution
comes from the $YM_2$ Wilson loop average. This will be done in the
following section.

\subsection{Heatkernel action from continuum limit}
 
We follow \cite{Mig2},sect.2.
 
Given a cubic lattice with spacing a we choose an axial gauge:
$\mbox{\bf A}_t = 0$. $Z_a (U)$ denotes the ``action'' associated with
the holonomy of an elementary cell.
 
The correlation function of a joint cell (see figure \ref{figure10})
 
 

\psbild{h}{figure10}{3cm}{}{joint cell}

will be given as
\begin{displaymath}
Z(U_1U_2) = \ \int dU Z_a(U_1 U^+)Z_a(U U_2)
\end{displaymath}
Generalising to a block of area A , decomposing $Z_a$ as a function of
the Wilson loop ($Z_a$ is a class function then) into a ``Fourier-''
series with a convenient choice of the ``Fourier-'' coefficient
and using the orthogonality of characters we have:
\begin{displaymath}
Z_A (V_C) = \  \sum_R (Z_{a,R})^{A/ a^2} dim(R) \chi_R (V_C)
\end{displaymath}
where $V_C$ is the holonomy of the block.$A/ a^2$is the number of
cells in  the block (see figure \ref{figure11}).
 
 

 
In the small area limit we have to have ($a \rightarrow 0 , A/ a^2
\rightarrow \infty $):
\begin{displaymath}
Z_{a,R} (V_C)\rightarrow 1 - \frac{1}{2} a^2 C_2 (R) \kappa^2
\quad \mbox{with} \quad V_C \rightarrow \mbox{\bf P} exp
\oint\limits_C \mbox{A}_\mu dx^\mu
\end{displaymath}
then we have
\begin{equation}
Z_A (V_C)=  \  \sum_R exp (-\frac{A \kappa^2}{2} C_2 (R) \ ) \  dim(R)
\ \chi_R (V_C)
\end{equation}
 Because this ``heat kernel action'' $Z_A$ is invariant under the
choice of the lattice spacing a it is the exact solution of $QCD_2$.
The action is ``additive'' because of the orthogonality of the
characters,
i.e. gluing two areas A, $A^{'}$ along an edge (like above) leads
to:
\begin{equation}
\int dU Z_A (V U^+)Z_{A^{'}} (U W) = Z_{A + A^{'}} (V W)
\end{equation}

The ``gluing'' property  and the diagrammatic representation of a
Riemann surface of genus G with b boundaries (see chapter 5) may be
used to evaluate the heat kernel action associated \cite{Cor1}
\begin{equation}
Z(A, G, U_1 .. U_b) =  \ \sum_R
exp(-\frac{A \kappa^2}{2} C_2 (R) \ ) \ dim(R)^{2 - 2 G - b}
\prod\limits_{i=1}^b \ \chi_R (U_i)
\end{equation}

\vspace{2cm}

\psbild{h}{figure11}{3cm}{}{block of area A}

\chapter{Some extracts: From group theory}
 
By making use of the ``Fourier-'' expansion of the partition function
the problem was solved in full generality. The solution, however, is
still quite formal: We just transferred the problem to the field of
group representations on a direct sum of irreducible invariant spaces.
 
We have to find explicit expressions for the Casimir eigenvalue {\em
on},
for the dimension {\em of} these spaces in terms of a complete, unique
classification, such that we can do mare than just writing down the
solution formally.
 
The expressions we find should be suitable for a 1/N-expansion.
 
The solutions shall be given, among some backround, below.

\newpage

\section{The (special) unitary groups}
 
The group structure of U(N), SU(N) is induced by their fundamental
representations as matrix groups on a N-dimensional vectorspace V over
complex field. These are defined as subgroups of the set of all
nonsingular matrices operating on this vectorspace (general linear
group) GL(N) by the following constraints.
\begin{displaymath}
u \in \mbox{U(N) or SU(N):} \quad u u^+ = u^+ u = 1
\end{displaymath}
\begin{displaymath}
u\in  \mbox{SU(N):} \qquad det(u)=1
\end{displaymath}
 
GL(N) is the largest of the matrix groups for given N. It has four
fundamental representations:
$\{ g\}$ acting on $V$,$\{ g^* \}$ acting on $V^*$ (complex conjugate,
$\{ \
g^t\}$ acting on
${\tilde V}$ (transposed, dual space), $\{ g^+\}$ acting on ${\tilde
V}^*$ (hermitean adjoint) (\cite{Tung} ch.13).
 
Because of the condition $u^+ = u^{-1}$ the number of independent
fundamental representations comes down to two for U(N). The
representations of SU(N) will be dealt with in the last section of
this chapter.
 
The obvious application of the definitions leads to representations of
the matrixgroups on tensor product spaces. By definition: any finite
dimentional representation is equivalent to a representation on a
subspace of $V^{\otimes m} \otimes V^{*\ \otimes n} $ for some $n,m
\in \mbox{\sc I}\!\mbox{\sc N}$.
 
The unitary groups are Lie-groups: They are groups with the features
of differentiable manifolds. In this sense they are compact. As
mentioned before this means that we have an invariant normalisable
(Haar-)measure on group space \cite{Nach} and any representation is
completely decomposable into finite dimensional representations.
 
Subspaces of a carrier space may be ``invariant'' under group
transformations: each element of this subspace is transformed into a
linear combination of vectors in this subspace regardless which
element is applied. A space which does not contain any nontrivial
invariant subspaces is ``irreducible'' and so is the associated
representation (``IRREP'').

For compact groups each IRREP is equivalent to a finite dimensional
unitary representation \cite{Serre} \cite{GTM98}.
 
\vspace{1em}
 
The symmetric group $S_n$ of permutations of (1..n) has
representations on $V^{\otimes n}$ by permutations of the indices of
the tensors in $V^{\otimes n}$. The IRREPs of $S_n$  are wellknown:
They are classified by Young tableaux (see eg. \cite{Tung}chapter5).
To a given Young tableau associated is the Young symmetriser $Y_n$,
which
generates the IRREPs of $S_n$ on $V^{\otimes n}$ (see figure
\ref{figure12}).


\psbild{h}{figure12}{3cm}{}{standard Young diagram}
 
First one has to sum over all permutations which leave the content if
the columns unchanged weighted by their sign (antisymmetrising the
columns, operator $a_{Y_n}$), then over all permutations of the rows
(symmetrising the rows, operator  $s_{Y_n}$). Then  $Y_n$ is given as:
 $Y_n= s_{Y_n} a_{Y_n}$.
 
There are different ways to describe  a Young tableau, each of which
gives a ``partition'' of n:
\begin{itemize}
\item by the number of boxes in each row:\ \ \   $\{ n_i \}, \ \sum
n_i = n\
$
\item by the number of boxes in each column: $\{ c_i \}, \ \sum c_i =
n$

\item by the number of columns of length i:\ \ $\{ l_i \}, \
\sum i l_i = n$
\end{itemize}
 
\newpage

By drawing a sketch it is easy to find:
\begin{equation}
c_k = \ \sum_i \Theta (n_i +1 -k) \qquad \Theta (n) = 1 \mbox{ if } n
> 0 , \ = 0 \mbox{ if } n \leq 0
\end{equation}
\begin{equation}
l_j = n_j - n_{j+1}
\end{equation}
All three ways will be used now and then in this work.
 
We will need the Young symmetriser for two things:
\begin{itemize}
\item the Young tableaux do not only classify the IRREPs of $S_n$,
they
classify the IRREPs of GL(N) on $V^{\otimes n}$, too. This is roughly
speaking, because the operations of $S_n$ commute with those of GL(N)
in the maximal possible way (\cite{Tung} chapter 5 and 13).
\item we will use the Young symmetriser to calculate the character of
transpositions $\chi_R (T)$ on an irreducible space R. By that we will
find a connection between the eigenvalue of the quadratic Casimir $C_2
(R)$ and $\chi_R (T)$, and at the same time we will solve the
``symmetry-factor'' of our string maps. This is, I would say, the most
crucial step in the interpretation as it will be outlined in this
work.
\end{itemize}
\vspace{1em}
Finally I want to state Schur's lemma:
 
An operator, which commutes with all group operations on an
irreducible space, is a multiple of the identity on this space (see eg
\cite{Tung} section 3.4)
 
This is needed at some stages of the considerations.

\section{On $C_2 (R)$}
\subsection{Lie algebras}
 
This section shall deal with the eigenvalue of the quadratic Casimir
$C_2 (R)$
and closely related problems. Because, as already stated, $C_2 (R)$
has ``more'' to do with the Lie algebra of the group than with the
group itself it is necessary to give some account of them first.
 
The elements of a Lie group are, in a neighbourhood of the identity at
least, given in terms of a finite number of real parameters:
As a Taylor expansion ($g \in G$)
\begin{displaymath}
g(\alpha ) = \ \mbox{\bf 1} \ + \sum_{k=1}^r \alpha_k (\frac{\partial
g}{\partial \alpha_k} )_{\alpha_k = 0} \ + O(\alpha^2)
\end{displaymath}
\begin{displaymath}
\equiv \ \mbox{\bf 1} \ + \sum_{k=1}^r \alpha_k X_k  \ + O(\alpha^2)
\end{displaymath}
 
The $ X_k$ are the generators of infinitesimal group transformations
and are a closed set under commutation
\begin{displaymath}
[  X_\sigma ,  X_\rho ]= c_{\sigma \rho}^\tau  X_\tau
\end{displaymath}
This defines the Lie algebra of the group.
 
The structure constants $c_{\sigma \rho}^\tau$ are antisymmetric in
the lower indices and transform as (2,1)-tensors under changes of the
Lie algebra basis.
 
The scalars
\begin{displaymath}
I_n = \ c_{\alpha_1 \beta_1}^{\beta_2} c_{\alpha_2 \beta_2}^{\beta_3}
 ... c_{\alpha_n \beta_n}^{\beta_1} X^{\alpha_1} ...  X^{\alpha_n}
\end{displaymath}
are invariant under group transformations. For a Lie algebra of rank l
there are l linearily independent such invariants (see eg. \cite{Corn}
\cite{Wyb}).
 
The operator $I_2 = C_2$ is called the quadratic Casimir and it is
quite easy to show that it satifies:
\begin{displaymath}
[C_2 , X_\rho] = 0 \quad \forall X_\rho \qquad \mbox{\cite{Wyb} 5.18}
\end{displaymath}
For arcwise connected Lie groups this is enough to show that the
Casimir is a multiple of the identity on an irreducible invariant
space $R$:
$C_2 |_R = C_2 (R) \mbox{\bf 1}_R$ , because then each element can be
expressed as pathordered exponential (see eg.\cite{Naka}p337).
 
U(N) is connected because each of its elements is diagonisable and its
exponential form can be found almost trivially (\cite{Fi} p202).
\footnote{ for othogonal groups the discrete qoutient group which
carries from one connected component to the others has diagonal
representations. I do not know whether this is always the case.
\cite{O'R}sect.1.2}
 
For semisimple Lie algebras like {\sc su}(N) the tensor $c_{\alpha
\beta}^{\gamma}c_{\gamma \delta}^{\alpha}$ is nondegenerate (Cartan's
criterium). This Killing form / Cartan metric then gives a scalar
product on the Lie algebra and the quadratic Casimir reads as $ T
\cdot T , \quad T=( X_1, .. X_r)$.

\subsection{Cartan-Weyl form}
 
For semisimple Lie algebras there exists a nice basis which makes it
easier to find the eigenvalues of $C_2$. In a semisimple Lie algebra
of
rank l there can be found, by definition, l operators in a Lie algebra
basis which mutually commute; they form the Cartan subalgebra (the
maximal torus) $\{ H_i \}$.
 
The remaining elements of the basis can be arranged in a set $\{
E_\alpha \}$ such that:
\begin{displaymath}
[ H_i , E_\alpha ] = \ \alpha_i E_\alpha
\end{displaymath}
The other relations for the Cartan Weyl form of semisimple Lie
algebras are:
\begin{displaymath}
[H_i , H_k] = 0 \qquad [E_\alpha , E_\beta ] = N_{\alpha \beta}
E_{\alpha + \beta}
\end{displaymath}
\begin{displaymath}
[E_\alpha, E_{- \alpha}] = \alpha^i H_i
\end{displaymath}
The vectors $E_\alpha$ are labeled by l numbers $ \alpha_i$. The
vector
$( \alpha_1, ..  \alpha_l) \equiv  \alpha$ lies in the ``root'' space.
Root vectors are called positive if the first nonzero component is
positive. A basis of the root space only containing positive roots,
consists of ``simple'' roots.
 
The indices in root space are raised and lowered by a metric induced
by the Killing form, $g_{\beta \delta}= c_{\alpha
\beta}^{\gamma}c_{\gamma \delta}^{\alpha} \ \ $, which is normalised
such
that:
$g_{\alpha, -\alpha} = 1$ i.e.:
\begin{displaymath}
g = \left( \begin{array}{cccccc}
     g_{i k} & & & & & \\
      & 0 & 1 & & & \\
      & 1 & 0 & & & \\
      & & & 0 & 1 & \\
      & & & 1 & 0 & \\
      & & & & & \ddots
           \end{array} \right)
\end{displaymath}
The matrix $g_{i k}$ is a metric on root space (\cite{Wyb} sect.6.5).
 
This form of a Lie algebra is the one usually used to
determine the spectrum of the angular momentum operators $J_z , \ J^2$
(\cite{Mess} 13.1):
\begin{displaymath}
{\vec J}= ( J_x , J_y , J_z ) \qquad J_+ = \frac{1}{\sqrt{2}}(J_x + i
J_y ) \qquad J_- = \frac{1}{\sqrt{2}}(J_x - i J_y )
\end{displaymath}
\begin{displaymath}
\Rightarrow [J_z , J_+] = J_+ \qquad [J_z , J_-] = -J_- \qquad [J_+ ,
J_-] =  J_z
\end{displaymath}
$J^2 = J_x^2 + J_y^2 + J_z^2$ is the quadratic Casimir of {\sc su}(2).
 
So we expect that it is possible to give an expression for the
eigenvalue of the Casimir in full generality in terms of a basis in
this form. Furthermore we expect a connection between the eigenvalues
of the Cartan subalgebra and the labeling of irreducible invariant
subspaces.
 
In this context the ``Cartan matrix'' will be used:
\begin{displaymath}
A_{i j} \equiv \frac{2 <  \alpha_i ,  \alpha_j >}{ <\alpha_i ,
\alpha_i >} \qquad \{  \alpha^i \} \mbox{ simple roots }
\end{displaymath}

\newpage

\subsection{Weights}
 
Since the l operators $H_i$ are mutually commuting a basis in each
representation may be found such that the  $H_i$ have the same
eigenvectors:
 $H_i | v > = h_i | v > $.
 
The vector $h = ( h_1 , .. h_l )$ is called the ``weight vector'' of
$| v_h >$. For $ E_\beta | v > \not= 0$ we have by the commutation
relations
\begin{displaymath}
H  E_\beta | v > = (h + \beta )  E_\beta | v >
\end{displaymath}
This establishes the connection between the root space of the algebra
and the eigenspaces of the algebra in representations.
 
Thereby can be proved that IRREPs can be labeled by ``highest
weights''
$\Lambda = \sum l_i \Lambda_i \ \ $, where the $ \{ \Lambda_i \} \ \ $
are the ``fundamental weights'' defined by
\begin{displaymath}
\Lambda_j = \ \sum_{k = 1}^l (A^{-1})_{k j} \alpha_k \qquad \alpha_k :
\mbox{simple roots} \quad  A : \mbox{Cartan matrix}
\end{displaymath}
and the $l_i$ are positive integers (\cite{Corn} sect.15.3).
 
In \cite{Corn}, sect.16.7 a discussion which applies to {\sc su}(N) is
given connecting the highest weight of an irreducible space with a
corresponding Young tableau:
 
$\Lambda = \sum l_i \Lambda_i \ \ $corresponds to a Young tableau with
$\{ l_i \}$ columns of length $\{ i \}$ ( \cite{Corn} p.650).
 
\vspace{1em}
Let $\delta$ be the sum over a half times the sum over all positive
roots, i.e.:
\begin{displaymath}
\delta = \frac{1}{2} \sum_{\alpha \in \Delta^+} \alpha
\end{displaymath}
Then the eigenvalue of the quadratic Casimir in an IRREP labeled by
$\Lambda$ is given as:
\begin{equation}
C_2 (R) = < \Lambda, \Lambda + 2\delta > \qquad \mbox{\cite{Wyb}15.1,
\cite{Corn}16.1}
\end{equation}

To find the eigenvalue $C_2 (R)$ in terms of the variables specifying
the Young tableau is now the problem to find a convenient form of the
algebra. This problem is solved for quite a while now. Some relevant
references and the result are given in the next section.

\subsection{The eigenvalue}
 
Some reference on the convenient choice of a basis for {\sc u}(N),
{\sc su}(N) was given already during the treatment of 't Hooft's U(N)
model. \cite{Corn}, App.F,G and  \cite{Gil} contain further reference.
 
The result of the calculation can be found in various references. The
formula used by D. Gross \cite{DG} \cite{GT1} can be found at
M.Resnikoff \cite{Res}, and very easily derived from the expression
given eg. in S.Okubo \cite{Oku}. It is:
\begin{displaymath}
C_2 (R) = \ N n + \sum_i n_i ( n_i + 1 - 2 i) \qquad \mbox{for {\sc
u}(N) }
\end{displaymath}
\begin{displaymath}
\equiv \ N n + {\tilde C} (R)
\end{displaymath}
\begin{equation}
 = \ N n + \sum_i n_i ( n_i + 1 - 2 i) - \frac{n^2}{N} \qquad
\mbox{for  {\sc su}(N)}
\end{equation}
where the notation of section 3.1 has been used; the last term in the
formula for {\sc su}(N) ``removes the trace''.
 
In \cite{DG} an explicit calculation is given establishing:
\begin{equation}
\sum_j c_j^2 = \ \sum_i n_i (2 i - 1) \qquad \mbox{using} \quad c_k =
\ \sum_i \Theta (n_i + 1 - k)
\end{equation}
leading to (for {\sc su}(N) )
\begin{equation}
C_2 (R) = \ N n \ + \sum_i ( n_i^2 - c_i^2) \ - \frac{n^2}{N}
\end{equation}
This means that in order to find the eigenvalue in terms of the column
we have to use \footnote{ Thereby the eigenvalue of the ``dual''
$R^*$, which is R with the rows and columns interchanged, has a
different eigenvalue for $C_2$. This should not be confused with the
eigenvalue of $C_2$ in its ``contragredient''
or conjugate representation ${\bar R}$ (cf.sect.3.4), which is the
same \cite{Rac}.}:
\begin{displaymath}
C_2 (R) = \ N n \ - \sum_i  c_i ( c_i + 1 - 2 i) \ - \frac{n^2}{N}
\end{displaymath}
So $ {\tilde C} (R)$ has the following forms:
\begin{equation}
{\tilde C} (R) = \sum_i n_i ( n_i + 1 - 2 i) =  -\sum_i  c_i ( c_i
 + 1 - 2 i) =  \sum_i ( n_i^2 - c_i^2)
\end{equation}

\section{On dim(R)}
 
The next quantity to be evaluated in terms of the Young tableau is the
dimension of the irreducible invariant subspace:$dim(R)$. This is done
by a formula given in \cite{Weyl}, p.201(5.14) \footnote{The
definition
of the difference product can be found at theorem 7.4.B, same reference}:
\begin{equation}
dim(R) = \frac{D ( \lambda_1, .. \lambda_N )}{D ( N-1, .. 1, 0 )} =
\prod_{i = 1}^N \prod_{j = i + 1}^N \frac{ \lambda_i - \lambda_j}{(N
-i)!}
\end{equation}
where $\lambda_i = n_i + N - i$.
 
A rather nice formula can be found for the relation to the
dimension of the associated representation of the symmetric group
$S_n$
given by \cite{Weyl}th.7.7p.213:
\begin{equation}
d_R = n! \frac{D( \lambda_1, .. \lambda_n) }{\lambda_1 ! .. \lambda_n
!}
\end{equation}
It can be found in \cite{Mur}, who give original reference for SU(N).
The
relation is for SU(N):
\begin{equation}
dim(R) = \ \frac{d_R}{n !} \prod_{i=1}^r \frac{\lambda_i !}{(N-i)!}
\end{equation}
where r is the number of not empty rows.
 
Using
\begin{displaymath}
\frac{(N + n_i - i) !}{(N-i)!} = \ \prod_{k = 1}^{n_i} (N + k -i) = \
N^{n_i} \prod_{k = 1}^{n_i} (1 + \frac{k - i}{N})
\end{displaymath}
one finds like in \cite{DG}:
\begin{equation}
dim(R) = \ \frac{d_R N^n}{n !} \prod_v (1+ \frac{\Delta_v}{N})
\end{equation}
where $v$ runs over all boxes in the diagram an $\Delta_v$ is the
difference of column and row index of a box.
 
\vspace{1em}
 
\newpage

By the ``hook'' formula \cite{GTM129},(4.12):
\begin{displaymath}
d_R = \ \frac{n !}{\prod_v (\mbox{hook  length})}
\end{displaymath}
(the ``hook'' length is the number of boxes right and beneath a box
plus one)
 
we see: $d_R = d_{R^*}$ and so:
\begin{displaymath}
dim(R^*)= \ \frac{d_R N^n}{n !} \prod_v (1- \frac{\Delta_v}{N})
\end{displaymath}
if one determines $\Delta_v$ from $R$ rather than from $R^*$, its dual
(rows and columns interchanged).

\section{On R}
 
For the ``Fourier-'' expansion we need all, but not more irreducible
representations.
In this section a way of labeling all representations in an
appropriate way is presented.
 
According \cite{Tung} the IRREPs of U(N) are given by a set containing
pairs of Young tableaux, one corresponding to representations on
$V^{\otimes n}$, the other on ${\tilde V}^{ \otimes m}$.
 
For SU(N) there exists an invariant  tensor, the Levi-Cevita symbol/
totally antisymmetric tensor with N indices: $\varepsilon$. Its
invariance stems from
the
condition $\mbox{det}(u) = 1$ and the connection between the
determinant and
the Levi-Cevita symbol. It may be used to raise/ lower indices. The
raised/ lowered form of an irreducible space will still be irreducible
because of the invariance of $\varepsilon$.
 
The application of $\varepsilon$ changes a representation $R$ into its
``conjugate'' ${\bar R}$: columns of length $c$ become columns of
length
$N - c$.

\newpage
 
Diagrammatically one gets the conjugate representation ${\bar R}$ of a
representation $R$ classified by rows $\{ n_i \}$ by drawing a
rectangle of $n_1 \times N$ boxes, drawing the columns of R from the
bottomline upwards preceeding from right to left (see figure \ref{figure13}).
 

 
${\bar R}$ is classified by columns $\{ {\bar c}_i \}$, rows $\{
{\bar n}_i \}$ given by:
\begin{equation} \label{contra}
{\bar c}_{n_1 - i + 1} = N - c_i \ , {\bar n}_i = - n_{N - i +1} + n_1
\quad \mbox{(see \cite{Tung}sect.13.4.2)}
\end{equation}

\vspace{1em}
 
So for SU(N) we have only one Young tableau for a unique
classification. Because the column of length $N$ gives the trivial
representation of SU(N) these Young tableaux may have column lengths
up to $N - 1$ to give a complete, and unique labeling of all IRREPs.
This is the result given in \cite{Weyl}\cite{Itz3}, and matches
perfectly well with the correspondence between highest weights and
Young tableaux (sect.3.2.3).

\psbild{h}{figure13}{6cm}{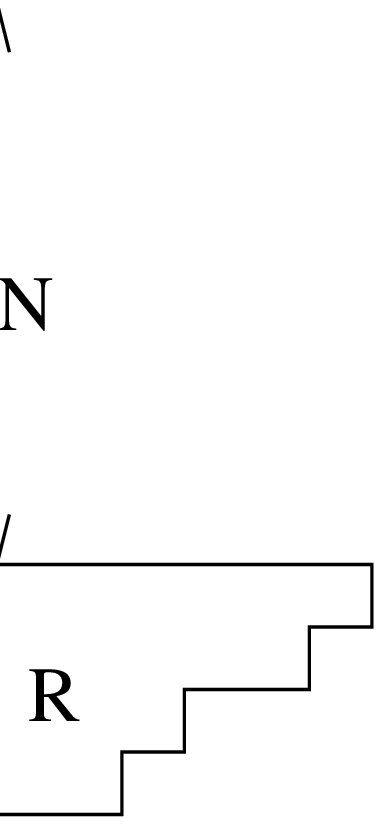}{conjugate representation}

\chapter{On the Wilson loop average}
 
Because, usually, Wilson loop averages are not among the objects of
physical education, I feel they should be given some 
consideration for themselves, though, of course, I can not give a
``complete''
treatment in any way. Now that the terms required for this are
introduced the discussion seems to be in order; along the way some
remarks of interest to related chapters will be made.
 
Section 1 will try to relate Wilson loop averages to some fields of
relevance to our dicussion. In section 2 the experimental proof
(Aharonov-Bohm effect), which is often said to have established the
importance of gauge potentials on their own (in Quantum Physics), is
viewed to prove the importance of Wilson loops. The final section will
be used to develop a Hamiltonian formalism on a Hilbert space directly
related to the Wilson loops.

\section{Gauge invariance and lattice theory}
 
In lattice gauge theories the Wilson loop average belongs to the
natural variables since holonomies are the essence of the theory. More
general: The gauge invariance as fundamental principle of the theory
means that its observables, gauge invariant quantities, can not depend
significantly on gauge dependent quantities like local values of a gauge
potential. The introduction of a gauge potential as a connection in
Yang-Mills theories means that physical varibles can not depend on the
local value of the gauge field: locally any such connection can be
trivialised , i.e. locally each gauge potential is in the same
conjugacy class as the identity (\cite{Naka}p.335). From this point of
view it is not counterintuitive that a set of nonlocal, gauge
invariant parameters could contain the physical information about a
system.
 
\vspace{1em}
 
Originally the concept of a loop average was introduced by F.Wegner
\cite{Wegner} connected with studies on Ising models with local $Z_2$
symmetry. It was adapted to nonabelian gauge theories by K.Wilson
\cite{Wils} in order to get a description of confinement.
 
Confined quarks are one of the ``phases'' of a quark system one could
think of. According to \cite{Itz1}(sect.6.1) any order parameter to
describe the different phases of a quark system should be nonlocal.
Such a quantity is the Wilson loop.
 
The string tension K, defined as
\begin{displaymath}
K \equiv \lim_{| C | \ large } -\frac{ln W (C)}{A (C)}
\end{displaymath}
where C is the loop, $| C |$ its length, A its enclosed area and $W
(C)$ the Wilson loop average,
\newline
enables to distinguish between confined ($K \neq 0$) and not confined
phase ($K = 0$).
 
Wilson comes about ``his'' loop when he considers a current-current
propagator:
\begin{displaymath}
D_{\mu \nu} (x) = < \Omega | T J_\mu (x) J_\nu (0) | \Omega >
\end{displaymath}
$ | \Omega >$ being the vacuum state, $J_\mu , J_\nu$ are thought to
correspond to $e^+ e^-$-annihilation with resulting quark/antiquark
pair production.
 
So it is diagrammaticly: (see figure \ref{figure28})
 
 
\psbild{h}{figure28}{4cm}{}{quark pair}

While calculating the amplitude in the Feynman picture one comes about
a weight factor $exp( i \kappa \oint A \ ds)$ (cf. section 2) and the
possibility of detecting a single quark corresponds to contributions
{}from loops with large $q {\bar q} $- separation. In the strong
coupling limit, when the area law holds, large separations are
suppressed.
 
\vspace{1em}
 
H.Rothe \cite{Ro} gives a teatment which suggests a simple relation
between the ground state energy of a particle/antiparticle pair and
the Wilson loop average. The discussion is given for QED (static
potential), though.
 
The two particle state is given as:
\begin{displaymath}
| \phi_{\alpha \beta} (x , y) > = {\bar \psi}_\alpha (x , 0) U(x , 0 ;
y , 0) \psi_\beta (y, 0) | \Omega >
\end{displaymath}
\begin{displaymath}
\mbox{where} \qquad U = exp( i \kappa
\int\limits_x^y dz^i A_i (z, t) )
\end{displaymath}
provides the right gauge transformation behaviour and the connection
to the Wilson loop.
The following relation arises when considering the time evolution of
the state:
\begin{displaymath}
T \rightarrow \infty : \quad < W_C [ A ] >\longrightarrow \quad
\sim exp( - E T )
\end{displaymath}
such that the ground state energy can be calculated as (up to an
additive constant):
\begin{displaymath}
E = - \lim_{T \rightarrow \infty} \frac{1}{T} \ \ln < W_C [ A ] >
\end{displaymath}

\section{Aharonov- Bohm effect}
 
The Aharonov-Bohm effect is the existence of an observable influence
of some current distribution on particles while there is no classical
force.
This effect was predicted by Y.Aharonov and D.Bohm \cite{YAB} and
observed by H.Boersch et al. \cite{Betal} and N.Osakabe et al.
\cite{Oetal}
in the situation depicted in figure \ref{figure30}. A jet of coherent
electrons
is splitted and guided around a flux tube through an area with no
classical field ($E, B = 0$), therefore there is no classical force on
the particles.
 
 
\psbild{h}{figure30}{7cm}{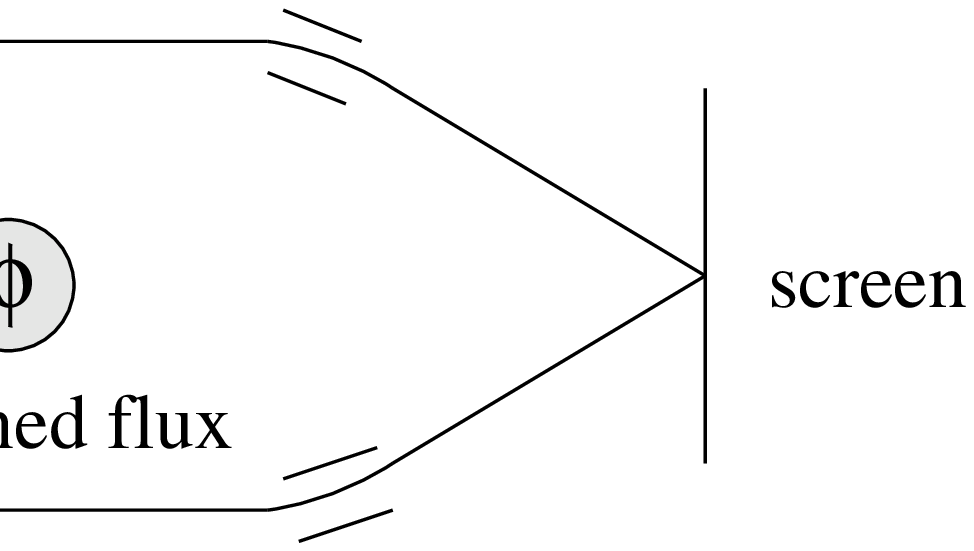}{experimental situation A-B effect}

Nevertheless the particles pick a phase factor due to the gauge field
in the area, which is not simply connected. There is no gauge where $A
= 0$ everywhere outside the flux tube. (see calculation below)
 
The experimental proof is then given by the shift of the interference
pattern due to a variation of the flux inside the tube.

The evolution of the state along the paths is given by (following
\cite{He}):
\begin{equation}
( i {\partial\!\!\!/ \ } - \kappa {A\!\!\!/ \ } - m) \psi \ = 0
\end{equation}
in the case above ($A_0 = 0$)
\begin{equation}
( i {\partial\!\!\!/ \ } - \kappa {\vec \gamma}{\vec A} - m) \psi \ =
0
\end{equation}
 
For each {\em one} path a gradient field ${\vec \nabla} f$ can be
found such
that
\begin{displaymath}
{\vec A} = {\vec \nabla} f \qquad f = \int\limits^{\vec r} d{\vec r}
\ {\vec A}
\end{displaymath}
 
By this gauge transformation
\begin{displaymath}
( i {\partial\!\!\!/ \ } - m ) \psi^{'} = 0 \qquad \psi^{'} = exp(- i
\kappa \int\limits^{\vec r} d{\vec r} \ {\vec A}) \psi
\end{displaymath}
 
Such that the part of the phase difference between the paths which
depends on the gauge field $A$ is:
\begin{equation}
\Delta \phi = - \kappa \oint d{\vec r} \ {\vec A}
\end{equation}
 
\vspace{1em}
 
So what is actually observed in the experiments {\em is } the ``Wilson
loop average''.

\section{A 2D Hamiltonian formalism}
 
The great success of field theories is their calculability by {\em
making} the description of interactions local using fields which
satify certain field equations. A theory based on nonlocal parameters
has to have a formalism for calculations. So to conclude this chapter
I want to give, following \cite{Cor1}, an example of an Hamiltonian
formalism based on Wilson loops.
 
\vspace{1em}
 
The easiest target space to consider is a cylinder: the periodicity in
the $x_1$ direction allows partial integrations without boundary
terms, because we require the field to be well defined. The straight
$x_0$- direction allows the choice of the axial gauge $A_0 = 0$.
 
We solve free gauge theory:
 
The equations of motion are \cite{Itz2}(p.568)
\begin{displaymath}
\partial^\mu F_{\mu \nu} + [ A^\mu ,  F_{\mu \nu}] = 0
\end{displaymath}
Implementing the axial gauge results in the necessity to demand
 the equation of motion resulting from the variation of $A_0$ as
operator equation on the space of physical states:
\newline
(in this gauge $F_{1 0} = \partial_0 A_1$)
\begin{displaymath}
\partial_1 F_{1 0} + [ A_1 ,  F_{1 0}] = 0
\end{displaymath}
 
Taking an infitesimal, time independent gauge variation $\delta_\omega$
\newline
(preserves $A_0 = 0$)
\begin{displaymath}
\delta_\omega \ A = \partial  \omega \  + [ A ,  \omega ]
\end{displaymath}
applied to a physical state $\psi$ leads to
\begin{displaymath}
\delta_\omega \ \psi (A) = \int dx \ (\delta_\omega \ A )
\frac{\delta}{\delta A} \ \psi (A)
\end{displaymath}
\begin{displaymath}
=\int dx  \  \omega^a \ ( - \partial \ \delta_{a b } \ + A^c f_{a b
c})
\ \frac{\delta}{\delta A_b} \ \psi (A)
\end{displaymath}
leads to the required operator equation:
\begin{displaymath}
(  \partial_1 \  \frac{\delta}{\delta A_1^a} \ + [
\frac{\delta}{\delta A_1} , A_1 ]_a ) \ \psi \ = \ 0
\end{displaymath}
Solutions of the form $ \psi [ \mbox{\bf P} exp \int_0^L dx_1 \ A_1]$
exist. Invariance under x- independent gauge transformations leads to
the conclusion that the Hilbertspace contains functions from
$ L^2 (G)$, the space of squared summable class function on the gauge
group G. The scalar product is given by the Haar measure as
\begin{displaymath}
< f_1 | f_2 > = \int dU \ f_1^* (U) f_2 (U)
\end{displaymath}
The Hilbertspace is separable by the characters of the group elements
in the finite dimensional IRREPs. $| U >$ denotes a state in $ L^2
(G)$ associated with U, $< R \ | \  U >$ is the trace of U in the
IRREP R.
Using the translational symmetry in $x_1$ we see that ``physical'' U's
are of the form:
\begin{displaymath}
U_\Gamma = \mbox{\bf P} exp \oint\limits_\Gamma ds \ A
\end{displaymath}
where $\Gamma$ is a circle at constant time $x_0$.
 
\vspace{1em}
 
Next we have to find the Hamiltonian. Usually $H = \frac{1}{2} ( E^2 +
B^2)$, but in two dimensions $B = 0$.
 
We have
\begin{displaymath}
\Pi_{A_1} \ = E \ = \frac{\delta}{\delta A_1}
\end{displaymath}
 
Taking an orthonormal basis for the Lie algebra $\{  T^a \}$,
$\frac{\delta}{\delta A_1^a}$ applied to a group element of the form
$U_\Gamma = \mbox{\bf P} exp \oint\limits_\Gamma ds \ A$ yields for
the trace $\chi_R (U_\Gamma )$
\begin{displaymath}
\Pi_{A_1^a} \ \chi_R (U_\Gamma ) \ =  \frac{\delta}{\delta A_1^a} \
\chi_R (U_\Gamma ) \ = \chi_R ( T^a \ U_\Gamma )
\end{displaymath}
(using the definition of the path ordered exponential and the rules
for fuctional derivatives)
 
Therby:
\begin{displaymath}
\frac{1}{2}\ (\Pi_{A_1^a})^2 \ \chi_R (U_\Gamma ) \ = \frac{1}{2}
\chi_R (
T^a T^a \
U_\Gamma ) \ = \frac{1}{2} C_2 (R)  \chi_R (U_\Gamma )
\end{displaymath}
\begin{displaymath}
\Rightarrow \quad H|_R =  \frac{1}{2} C_2 (R)
\end{displaymath}
 
So the quantum mechanical propagator for two states $| U_1 > , \ | U_2
>$ (both at $t = 0$) is given by: (cf. figure \ref{figure29})
 
 
\psbild{h}{figure29}{5cm}{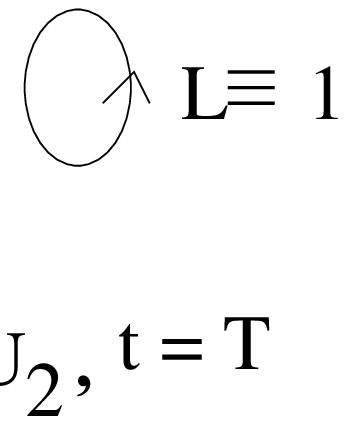}{quantum mechanical propagator}

\begin{displaymath}
Z (  U_1,  U_2, T ) = <  U_1 | \ e^{- T H } \  | U_2 >
\end{displaymath}
\begin{displaymath}
= <  U_1 | \sum_R | R ><  R | \ e^{-T H } \  \sum_{R^{'}} | R^{'}
><R^{'}
 | | U_2 >
\end{displaymath}
\begin{displaymath}
= \ \sum_R <  U_1 | R > \ e^{-  \frac{1}{2} T \ C_2 (R) } \ <  R | U_2
>
\end{displaymath}
\begin{displaymath}
= \ \sum_R \chi_R (U_1^+ ) \ \chi_R (U_2 ) \  exp( -  \frac{1}{2} T \
C_2 (R) \ )
\end{displaymath}
 
\vspace{1em}
 
The possibility to make an infinitesimal region flat means that the
holonomies around such a region are the identity then and the space of
states on such a target space has the form:
\begin{displaymath}
\psi (U) = \delta (U)
\end{displaymath}
 
Using both results and gluing an infinitesimal disc to the cylinder we
get for a plaquette of area A:
\begin{displaymath}
Z (A , U ) =  \ \sum_R \ dim(R) \ \chi_R (U ) \ exp( -
\frac{1}{2} A \ C_2 (R) )
\end{displaymath}
 
J.Baez and W.Taylor \cite{Baez} develop a Hamiltonian formalism using
``loop variables'' (cf. chapter 8).

\chapter{Some extracts:  About Riemann Surfaces}
 
By reducing a geometrical object to a topologically equivalent
triangulation it is possible to transfer the problem of finding
topological characteristics of this object to a field where algebraic
methods, from group theory in particular, can be applied.
 
Homology, homotopy and monodromy groups are such methods. Their concept
is to assign elements $ \{ g_i \}$ of  known groups (here: {\bf Z },
and $S_n$ respectively) to geometrically ``distinct''
objects $ \{ x_i \}$ . A group structure is induced for these objects
by their partners: $ x_i^{g_j}x_i^{g_k} =  x_i^{g_j g_k}$. In this
sense the group structure is ``formal''.
 
The ``geometrically distinct'' objects are found when one specifies
subsets of the triangulations and, possibly, identifies their elements
by equivalence relations. The representatives of such equivalence
classes (``generators'') are then the objects that recieve a partner
{}from the group.
 
In the cases under consideration the number of generators is always
finite and the structure of the groups is quite simple.
 
In these concepts do fit the ``branched coverings'', too. They are
introduced to have strings in the theory. Their classifying group, the
monodromy group, is a subgroup of $S_n$ and establishes the connection
to the theory of the gauge group.
 
\section{Homology, Homotopy}
 
Given a set of ``generators'' $ \{ x_i \}$ we introduce a group
structure by  $ x_i^{g_j}x_i^{g_k} =  x_i^{g_j g_k}$ (we take $g_j \in
\mbox{\bf Z}$).
 
Because a priori nothing is known about a possible Abelian character
we have to assume that every element in this finitely generated group
can be written as a ``word'' using the generators as ``letters'':
\begin{displaymath}
w = \ x_{j_1}^{i_1} ... x_{j_n}^{i_n}
\end{displaymath}
for some finite positive number n.
 
A word is called ``reduced'' if $\forall j \ : i_j \not= 0 , \ x_j
\not= x_{j+1}$; the empty word ($\forall j \ : i_j = 0$) is {\bf 1}.

The product of two words is obviously:
\begin{displaymath}
v w = \ \{x_{j_1}^{i_1} ... x_{j_n}^{i_n} \} \{ x_{l_1}^{k_1} ...
x_{l_m}^{k_m}\} = \ x_{j_1}^{i_1} ... x_{j_n}^{i_n}  x_{l_1}^{k_1} ...
x_{l_m}^{k_m}
\end{displaymath}
 A group G generated is this way with no more constraints is ``free''.
Abelian groups satisfy:
\begin{displaymath}
a_i a_j =  a_j a_i \Leftrightarrow a_i a_j
a_i^{-1} a_j
^{-1} = \mbox{\bf 1} \quad \forall i , j
\end{displaymath}
 and words in those groups can be furtherly reduced using these
relations. \newline (The homology group is abelian)
 
So in general a finitely generated group is defined by a set of
generators  $ \{ x_i \}$, the inducing group and a set of relations  $
\{ r_j \}$, under which words are identified (i.e. reduced to the same
reduced form). The group then consists of these reduced words. But for
our purpose the form $G = ( \{ x_i \}$,  $\{ r_j \}$) is the
appropriate
one (\cite{Naka}sect.4.4).
 
On homology: \newline
Taking a triangulation K of a surface M one has a lot of triangles
(equally well polygons). One may build surfaces or lines with them by
taking connected subsets of the triangulation.
Taking connected edges of the triangles we get ``1-chains'', the
1-chains without ends are called ``1-cycles''. They generate the
1-cycle group $Z_1 (K)$.
 
The boundaries of connected subsurfaces of the triangulation generate
the the 1-boundary group $B_1 (K)$.
 
Identifying all 1-cycles which together are a boundary gives the
first homology group
\begin{displaymath}
H_1 (K) = \ Z_1 (K) /  B_1 (K)
\end{displaymath}
 
The homology groups of two triangulations $ K , K^{'}$ of the same
surface M are isomorphic; the homology group of M may be defined and
taken to be the one of some triangulation K:
\begin{displaymath}
H_1 (M) \  \equiv \ H_1 (K) \  \cong \ H_1 (K^{'}) \quad
\mbox{\cite{Naka}def.3.12,13,16}
\end{displaymath}
 
For compact orientable surfaces of genus g one can choose 2g
generators in the way sketched in figure \ref{figure14}.

\vspace{1cm}


\psbild{h}{figure14}{8cm}{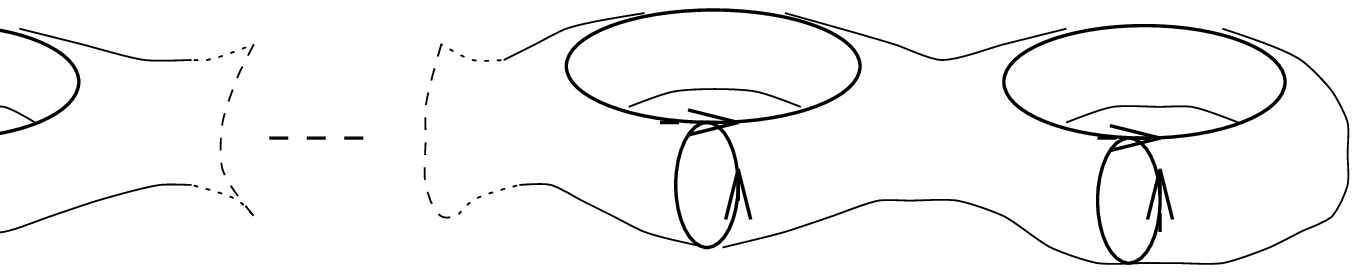}{homology generators}

\newpage
 
After deformation of the surface the generators may be chosen such
that they all have a common basepoint P. The generators have an
orintation. ``Cutting'' along the generators and identifying according
to the orientation of corresponding edges leads to a representation of
a 2g-surface as a 4g-gon (\cite{Naka}fig.2.6,12) (see figure \ref{figure15},
where the assigned group elements stand for the way one has to travel
over the generators for a closed path along the 4g-gon).


\psbild{h}{figure15}{4cm}{}{representation as a 4g-gon}
 
We leave the treatment of boundaries to the end of the paragraph on
homotopy.
 
\vspace{1em}
On homotopy: \newline
The objects leading to the homotopy group are ``loops''; these are
continous ``paths''
$\alpha : [ 0 , 1 ] \rightarrow M $
with $\alpha ( 0 ) = \alpha ( 1 )$.
 
The product of paths with $\alpha ( 1) =  \beta ( 0 ) $ is $\alpha *
\beta :\
  [ 0 , 1 ] \rightarrow M$
\begin{displaymath}
\alpha * \beta ( s ) = \alpha ( 2s) \quad \mbox{if } \ s \in  [ 0 ,
\frac{1}{2} ] \quad , \quad = \beta ( 2s -1 ) \quad \mbox{if } \ s \in [
\frac{1}{2} , 1 ]
\end{displaymath}
The same definition applies to loops.

Two loops based at $x_0 \in M$ are said to be equivalent
(``homotopic'')
if there exists a continuous map $F:  [ 0 , 1 ] \times  [ 0 , 1 ]
\rightarrow M$ with
\begin{displaymath}
F ( s , 0 ) =  \alpha ( s ) \qquad F ( s , 1 ) = \beta ( s ) \qquad F
( 0 , t ) = F ( 1 , t ) = x_0 \quad \forall t \in  [ 0 , 1 ]
\end{displaymath}
 
The product and the equivalence relation ``homotopy'' define the first
homotopy group $\Pi_1 ( M )$ (``fundamental group'')
(\cite{Naka}def.4.6,lemma4.7,th.4.8)
 
The generators given for the homology group of the torodi of genus g
generate the homotopy group, too. In terms of these generators:
\begin{displaymath}
\Pi_1 ( M ) = \ ( \{ a_i , b_i \}_{i \in g} ; \prod_i  a_i b_i
a_i^{-1} b_i^{-1} = \mbox{{\bf 1}} ) \qquad \mbox{\cite{Naka}(4.26)}
\end{displaymath}
 
The classification of torodi described above is equivalent to the
classification of a genus g surface as a two sphere with g handles
\cite{Still}1.3.7. In both formalisms orientable surfaces with m
boundaries
are included by making m punctures in the sphere, the 4g-gon
respectively.

The new generators are included by sending loops from P around the
punctures(\cite{Still}1.3.9). Viewed on the punctured sphere the
formalism is convincing
and the homotopy group is given by:
\begin{displaymath}
\Pi_1 ( M \setminus \{ q_j \}_{j \in b}) =  \ (  \{ a_i , b_i ,  c_j
\}_{i
\in g , j \in b } ; \prod_j  c_j  \prod_i  a_i b_i
a_i^{-1} b_i^{-1} = \mbox{{\bf 1}} )
\end{displaymath}
 

\psbild{h}{figure16}{3.5cm}{}{representation of b boundaries}
 
\section{Branched coverings}
 
The mathematical interest in Riemann surfaces stems from the fact that
the concept of a ``complex structure'' like in the complex plane can
be introduced on them. Because of this Riemann surfaces are used to do
complex analysis and a lot of treatments on Riemann surfaces involve
heavy complex analysis.
 
The concept of ``complex structure'' on manifolds is local: basicly
only an operator $J$ with $J^2 = - 1$ has to exist on the manifold
(\cite{Naka}p.275). But the full strength of complex analysis can only
be achieved on manifolds which admit such a structure globally. These
manifolds are ``complex'', i.e. they can be given by local charts
$\phi : M \rightarrow \mbox{\bf C}$.
 
The concept of ``holomorphicity'' on {\bf C} is local: the
Cauchy-Riemann equations are local. This term is introduced for
maps between complex manifolds M, N (Riemann surfaces are complex):
functions $f : M \rightarrow N$ are holomorphic, if, in terms of local
charts $\phi ( \psi )$ on M (N),  $\psi \circ f \circ \phi $ is
holomorphic in {\bf C}.
 
After the choice of a specific complex structure on N there 
is only one complex
structure on M to make a differentible map $f : M \rightarrow N$
holomorphic:
The pullback $f^* ( J )$ \cite{Ahl}.
 
Because the arrangements described above prove to be useful Riemann
surfaces
are defined to be one (complex) dimensional analytic manifolds
(\cite{Fark}p.9). Usually the requirement of connectedness is added.
For the results needed in our context this restriction is not
necessary and, because a partition function has something to do with
diconnected diagrams, not desirable either. The mathematical
statements needed and given below generalise to disconnected Riemann
surfaces.
 
Holomorphic maps f between Riemann surfaces M, N have nice features,
eg. $f : M \rightarrow N$ holomorphic, M compact, then f is either
constant or surjective. In the latter case N is also compact
(\cite{Fark}I.1.6).
 
``Covers'' are locally diffeomorphic, surjective mappings between
manifolds (\cite{Naka}def.4.48).

Such surjective maps have definite ``winding numbers'':
A nonconstant holomorphic cover $f : M \rightarrow N$ takes on the
value $f ( P )$ n times ``for all'' $x \in M$. (for some finite
positive number n) (\cite{Fark}I.1.6).
 
Consider $N = \mbox{\bf C} \cup \{ \infty \}$, the Riemann sphere, and
the map $w = z^2$. w runs runs two times around {\bf C}; the map
$\sqrt{w}$
covers {\bf C} twice; there are two fixed points: $0 , \infty$
(``branch'' points).
 
w may be cut into two ``sheets'' along a great circle between the two
branch points $0$ and $\infty$. The edges of the sheets are identified
such that the result is w again. The covering from w onto {\bf C} is
then locally diffeomorphic.
 
Inverse images of paths around or through 0 (or $\infty$)(under the
covering map) may start on one sheet and end after identification at
one of the edges on another: branch points are, in this sense,
``tubes'' lifting from one sheet to another.
 
In the neighbourhood of the branch points 0 and $\infty$ the covering
space w looks in
suitable local coordinates like $z^2$; two is here the number of
sheets.
\newline (example from \cite{Still}1.1.1)

All these statements generalise for coverings M (Riemann surface) of a
Riemann surface N: $f : M \rightarrow N$ a holomorphic cover.
 
f has a finite number of ``branch'' points P on N (the branch locus S)
in whose neighbourhoods there exist local coordinates such that the
covering locally looks like $z^n$, where n is constant everywhere and
is the winding number. Each branch point gets a branch index $b ( P )
= n - 1$. The branching number B is defined to be
\begin{displaymath}
\sum_{P \in S} b ( P ) \equiv B
\end{displaymath}
The definitions are closely related to the Riemann-Hurwitz relation
and its derivation
 
The Euler-Poincar${\acute e}$ characteristic for compact surfaces is
given as
\begin{displaymath}
\chi = V - E + F \qquad \mbox{(see chapter 1)}
\end{displaymath}
 For orientable
surfaces:$\chi = 2 - 2 g$, where $g$ is the genus. The
Euler-Poincar${\acute e}$ characteristic has the same definition for
connected as well as for disconnected surfaces and may serve as the
defining quantity for the ``more general'' genus using the relation
above. The genus $g$ of such disconnected sets of surfaces may have
any value between $- \infty$ and $+ \infty$.
 
\newpage

The Euler-Poincar${\acute e}$ characteristic $\chi$ is all that is needed
to prove the  Riemann-Hurwitz relation (\cite{Fark}I.2.6,I.2.7):

M, N Riemann surfaces of genus g, G; $f : M \rightarrow N$ a
holomorphic map of degree n, branching number B; then
\begin{equation}
2 ( g - 1 ) = 2 n ( G - 1 ) + B \qquad \mbox{remark: B is always even}
\end{equation}
 
Closely related is Kneser's formula, which is not needed in the
following treatment (\cite{Zie}p.73):
\begin{equation}
2 ( g - 1 ) \geq 2 n ( G - 1 )
\end{equation}

\section{Monodromy group}
 
As in the example of section 5.2 we imagine the covering space M to be
cut along great circles between the branch points into n sheets. They
are labeled from 1 to n and the edges are identified; because outside
the branch points the covering map is locally diffeomorphic, and even
though partially disconnected, the sheets of one {\em connected}
component of
the covering have , after the identifications,  to be {\em connected}
again.
 
An identification is a permutation $\Pi_q$ of $( 1 ... n )$, which tells
where a circle around the branch point q, starting after one
identification in the sheet i, carries with the next identification:
onto the sheet  $\Pi_q ( i )$.
 
These permutations, associated to the branch points, form a subgroup
of the symmetric group, the ``monodromy group'' (\cite{Still}p.57).
Considering branched coverings of the sphere with m branchpoints one
finds the relation of the monodromy group.
 
Choose a point P on some sheet i outside every branch point, cut the
sheets on all levels along great circles from P to every branch point.
A circle around P (see figure \ref{figure17}) ends on the sheet
\begin{displaymath}
\Pi_m \Pi_{m - 1} ... \Pi_2 \Pi_1 ( i )
\end{displaymath}
but also has to end on the sheet i , because P is not a branch point.
So we have:
\begin{equation}
\Pi_m \Pi_{m - 1} ... \Pi_2 \Pi_1 \ = \ \mbox{\bf 1}
\end{equation}
This looks exactly like the relation defining the fundamental group
$\Pi_1 ( S^2 \setminus S)$.


\psbild{h}{figure17}{3.5cm}{}{consistency relation}

And, indeed, the definition above induces an homomorphism
\begin{displaymath}
\Pi_1 ( S^2
\setminus S , P) \rightarrow S_n
\end{displaymath}
and furthermore: the treatment
generalises by the same procedure, only slightly more complicated by
the identifications along the generators for the holonomy group, for
any Riemann surface, such that an homomorphism: $\Pi_1 ( N
\setminus S , P) \rightarrow S_n$ is given. (P is the basepoint of the
groups.)
 
There exists a theorem which makes these homomorphisms most relevant
for our discussion. For this first two definitions:
\begin{itemize}
\item Two branched covers $f_1 , \ f_2 : M \rightarrow N $ are said to
be {\em equivalent} if there exists diffeomorphism $\phi : M
\rightarrow M$ such that $f_1 \circ \phi = f_2$
\item Two homomorphisms $\psi_1 , \psi_2 : \Pi_1 ( N
\setminus S , P) \rightarrow  S_n$ are said to be  {\em equivalent} if
they differ only by an inner automorphism (``similarity
transformation'') of $ S_n$, i.e.
$\exists g \in S_n : \psi_1 = g  \psi_2 g^{-1}$
\end{itemize}
 
The theorem is \cite{Eze} \cite{Fult} \cite{Cor1}:
\newline
S a finite subset of a Riemann surface N, n a positive integer, $P \in
N , \ P \notin S$. \newline
There is a one to one correspondence between equivalence classes of
homomorphisms $ \psi : \Pi_1 ( N\setminus S , P) \rightarrow  S_n$ and
equivalence classes of n-fold branched coverings with all elements of
S as branch points.
 
\vspace{1em}
 
This theorem will be used to calculate a ``symmetryfactor'' $| S_n |$
connected with a branched covering, which is already determined by the
$QCD_2$ partition function, but can be reproduced consistently by this
theorem within the string picture.\footnote{The argumentation in
\cite{GT1} procedes formally a little bit differently; the connection
is given in \cite{Cor1}sect.5.1.2}

\chapter{Interpretation:  One chiral sector}
\section{Introduction}

A QFT-perturbation theory is usually a set of Feynman-rules which
define a connection between a set of topological objects, the
diagrams, and probability amplitudes for physical processes.
 
The Feynman diagrams are ``topologically inequivalent'', i.e. certain
geometrical objects, and associating things like vertices, propagators
etc. means that they are embedded in spacetime; only the topological
features of the embedded form are relevant to the theory.
 
String perturbation theory, in these respects, follows the same
procedure.
``Embedding'' strings in a two dimensional compact manifold makes
things special. To make these maps surjective they only have to be
holomorphic.
A consistent interpretation using branched coverings proves to be
possible.
 
Orientations, which one may assign to orientable diagrams, strings or
the skeletons of common QFT, have no meaning to the amplitudes of free
gauge theory. Meaningful orientations are only associated to fermion
lines, which are, following 't Hooft's work, considered to be
suppressed in $N \rightarrow \infty $.
 
In the perturbative expansion of the partition function related to
actions such as the Nambu-Goto action, one could equally well sum over
both orientations and all strings, if these admit orientations: This
would only result in a new normalisation of the partition function.
 
In chapter 7 and 8 some argumentation will be given for introducing
two sets of strings with ``opposite'' orientation. In this context the
discussion given in this chapter applies to each {\em one} set of
surfaces. Therefore the term ``chiral'' partition function or
``chiral'' sum.
 
The partition function contains four terms which have to be understood
in terms of strings:
\begin{displaymath}
Z ( G , \lambda A , N) = \sum_R dim(R)^{2 - 2 G}exp(-\frac{\lambda A
C_2 (R)}{2 N})
\end{displaymath}
\begin{displaymath}
 = \sum_R dim(R)^{2 - 2 G} exp(-\frac{\lambda A
n}{2})
exp(-\frac{\lambda A}{2 N} {\tilde C} (R) ) exp( \frac{\lambda A
n^2}{2 N^2})
\end{displaymath}
where $\lambda = \kappa^2 N$ chosen to take the limit motivated by 't Hooft
\cite{tH1} $N \rightarrow \infty \ \wedge \ \lambda = \kappa^2 N = const.$
(cf. section 1.2.2).

The terms of the 1/N-expansion related to the first two exponentials
will be understood by means of branched coverings. The first one
already looks like the area of a string which winds n times around the
target space of area A (sections 3 \& 4).
 
The last term give rise to the introduction of ``tubes'' and
``collapsed'' (``infinitesimal'') handles (section 5).

\newpage

\section{Chiral 1/N- expansion}
 
In the recent chapters the following formulae were achieved:
\begin{displaymath}
Z = \sum_R dim(R)^{2 - 2 G} exp(-\frac{\lambda A}{2 N} C_2 (R) )
\end{displaymath}
\begin{displaymath}
dim(R) = \ \frac{d_R N^n}{n !} \prod_v (1+ \frac{\Delta_v}{N}) \
\approx  \frac{d_R N^n}{n !} \ + O(N^{n - 1})
\end{displaymath}
\begin{displaymath}
C_2 (R) = \ N n \ + {\tilde C} (R)\ - \frac{n^2}{N}
\end{displaymath}
Using these and following the discussion of the previous section
leads to the expansion below:
\begin{displaymath}
Z = \sum_{n = 0}^\infty \sum_{R \in Y_n}^{'} dim(R)^{2 - 2G}
exp(-\frac{\lambda A}{2 N} C_2 (R) )
\end{displaymath}
\begin{displaymath}
\approx \sum_{n = 0}^\infty \sum_{R \in Y_n}^{'} exp(-\frac{\lambda A
n}{2}) \quad \times
\end{displaymath}
\begin{equation}
\sum_{i = 0}^\infty ( \frac{(- \lambda A {\tilde C} (R))^i}{2^i i !}
N^{n ( 2 - 2 G ) - i} + O( N^{n ( 2 - 2 G ) - i - 1}))
\end{equation}
The prime on the summation indicates that only the Young tableaux with
columns of length less or equal $N - 1$ are to be summed over.
 
The O(..) indicates the subleading contributions from the last term in
$C_2 (R)$ as well as contributions from the dimension (for $G \neq
1$).
 
The branching number in the Riemann-Hurwitz relation, whose
look-a-like
can be found above, is always even, while the summation index i may
take any value in the limits. But although the Riemann-Hurwitz
relation applies only to half of the sum we have no serious problem
with that: The formalism will be generalised such that we can
include odd ``branching'' numbers  naturally.

\section{Strings and branched coverings}
 
Taking $\lambda$ as the string tension and 1/N as string coupling we
know quite a lot about a string expansion of closed, orientable
strings:
 
 $\lambda$ should always appear together with the area A and a string
of genus g contains $2 g - 2$ vertices, so have to have a factor $N^{2
- 2 g }$ in the expansion (see figure \ref{figure18}).
 

\psbild{h}{figure18}{8cm}{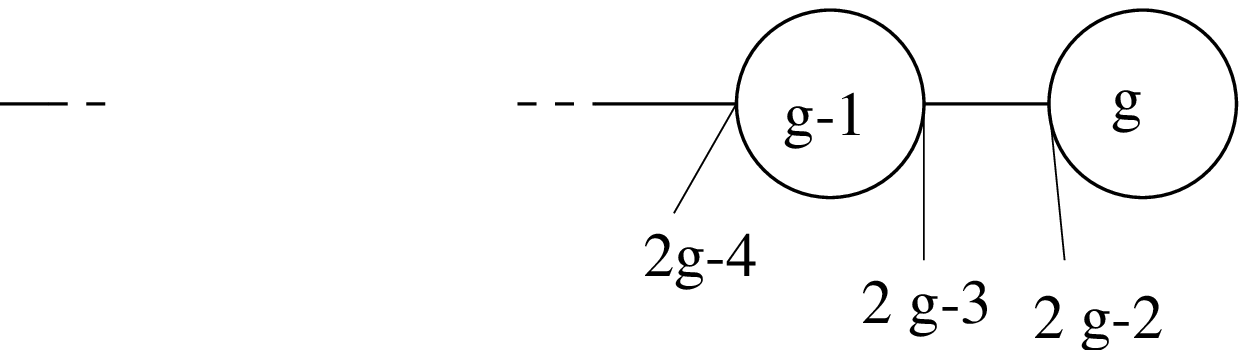}{holes and vertices}
 
Because the perturbation theory of a partition function contains
disconnected diagrams, the series in g has to range from $- \infty$ to
$+ \infty$ (with $\chi \equiv 2 - 2 g$) (cf.section 5.2).
 
Observing the Riemann-Hurwitz look-a-like in the expansion we are
tempted to write down:
\begin{displaymath}
Z( G , \lambda A , N ) = \sum_{g = - \infty}^{+ \infty} \sum_n \sum_i
\zeta_{g , G}^{n i} exp( -\frac{n \lambda A}{2}) (\lambda A)^i N^{2 -
2 g}
\end{displaymath}
\begin{displaymath}
\mbox{with} \qquad \zeta_{g , G}^{n i} = \sum_R (\frac{n !}{d_R})^{2 G
- 2} \frac{1}{i !} (\frac{{\tilde C} (R)}{2})^i
\end{displaymath}
In the interpretation n will be the winding number and i will be the
``branching'' number of a covering.
 
As usual in perturbation theory there will be a ``symmetry factor''
associated with a diagram. This will turn out to be essentially the
coefficient $\zeta_{g , G}^{n i}$: $i ! \zeta_{g , G}^{n i} =
\mbox{symmetry factor}$.
 
It would seem to be intuitive to weight the coverings with winding
number n and ``branching'' number i, which are the labels we are
summing over, with the number of {\em different}
coverings with these coefficients. This approach will give the
required term for the interpretation.
 
Usually symmetry factors are determined such that the theory becomes
unitarian. Because $QCD_2$ is already a gauge field theory reproducing
the right terms by defining a symmetry factor should yield a correct
``string'' theory.

\section{Calculating the symmetryfactor}
 
For the calculation of the symmetry factor we will count, as described
in chapter 5, the number of distinct homomorphisms $\# \{ [ H_\nu ] \}$ :
\begin{displaymath}
H_\nu: \Pi_1 (N \setminus \{ q_i \} ) \rightarrow S_n
\end{displaymath}
 connected with
a covering map $\nu$. The set of covering maps of a target space with
genus G with winding number n and ``branching'' number i will be
denoted as $\Sigma (G , n , i )$.
 
For the purposes of counting the distinct homomorphisms the branch
points with branching index $b ( q_i ) = n - 1$ may be decomposed into
$n - 1$,
what I want to call, ``decomposition'' points:
 
 To each branch point
$q_j$ is associated a permutation $\Pi_j$ of $( 1 ... n)$. Now: every
permutation may be decomposed into at most $n - 1$ transpositions. It
is possible, obviously, to define some algorithm such that the
decomposition becomes unique.
 
The permutation is then a sequence of transpositions. When the
homomorphism was defined all that was required was to know where a
path starting on some sheet i, going around some branch point is
heading with the next identification. This involves one of the
transpositions only. So we may associate to each transposition one
point, which connects two sheets only and may be located anywhere on
the targetspace. In spite of the similarity to branch points I will
call these points ``decomposition'' points to mark their origin and
their character.
 
This decomposition is helpful to calculate the symmetry factor and,
thereby, to find the correspondence to the chiral sum, and it is
necessary to generalise the formalism to odd ``branching'' numbers.
 
The decomposition points may be distributed in $\frac{(\lambda A)^i}{i
!}$ ways: each decomposition point has the whole area of the target
space as possible location, but the labeling of the points is
unimportant. This gives already one of the required terms.
 
\vspace{1em}
 
If one sums over i transpositions and $2 G$ general permutations with
a deltafunction sensitive to the defining relation of the monodromy
group, which is invariant under conjugation of its argument, one will
get
$n !$ the number of distinct homomorphisms, because of the freedom of
conjugation. Summing with a constant weight $\frac{1}{n !}$ should
give the right result. the actual evaluation of this sum has, of
course, to do with the representation theory of the symmetric group.
 
\vspace{1em}
 
For the symmetric group the following equations are valid:
\begin{equation} \label{a}
\frac{1}{n !} \sum_R \chi_R ( \mbox{{\bf 1}} ) \chi_R ( \rho ) =
\delta ( \rho ) \qquad \mbox{\cite{Tung} (3.6-3)}
\end{equation}
\begin{equation} \label{b}
 \sum_R  \chi_R ( \rho ) D_R (\rho^{-1} ) = \frac{n !}{d_R} \mbox{{\bf
1}}_R \qquad
\mbox{\cite{Tung} th.3.5}
\end{equation}
\begin{equation} \label{c}
\sum_{\sigma \in S_n}  D_R (\sigma \rho \sigma^{-1} ) =  \frac{n
!}{d_R} \chi_R ( \rho ) \mbox{{\bf
1}}_R \qquad
\mbox{\cite{Tung} cf. proof of lemma 3.7}
\end{equation}
\begin{equation} \label{d}
\sum_{\tau \in T} D_R (\tau ) = \frac{1}{d_R} \frac{n (n - 1)}{2}
\chi_R ( \tau )
\mbox{{\bf 1}}_R \qquad \mbox{\cite{Tung}lemma to th.3.7}
\end{equation}
where T is the conjugacy class of transpositions and $ D_R (\tau )$ is
the representative of $\tau$.
 
The sum described before reads as
\begin{equation}
\sum_{\tau_1 .. \tau_i \in T} \sum_{s_1 , t_1 .. s_G , t_G \in S_n}
\frac{1}{n !} \delta (\tau_1 .. \tau_i \prod_{j = 1}^G s_j t_j
s_j^{-1} t_j^{-1} )
\end{equation}
{}From (\ref{c}) we have:
\begin{equation}\label{e}
\sum_{s , t \in S_n} D_R (s t s^{-1} t^{-1} ) = (\frac{n !}{d_R})^2
 \ \mbox{{\bf 1}}_R
\end{equation}
 
Rewriting the delta function with (\ref{a}), writing the character as
trace over a product of matrices and using (\ref{d}) and (\ref{e})
leads to:
\begin{displaymath}
\sum_R \ (\frac{1}{n !})^2 \ d_R \ (\frac{n !}{d_R})^{2 G} (\frac{n
(n-1) \chi_R ( T )
}{2 d_R})^i \ tr  \mbox{{\bf 1}}_R
\end{displaymath}
\begin{displaymath}
= \sum_R \  (\frac{n !}{d_R})^{2 G - 2} \  (\frac{n (n-1)
\chi_R ( T ) }{2 d_R})^i
\end{displaymath}
 
What remains to be done is the calculation of the character of
transpositions in an arbitrary IRREP R. One possible procedure is the
following:
 
The operator
\begin{displaymath}
{\tilde T} = \sum_{\tau \in T} \tau
\end{displaymath}
commutes with $S_n$ by the rearragement lemma, since every permutation
is decomposable into transpositions. According to Schur's lemma
${\tilde T}$ has to be a multiple of the identity on any IRREP of
$S_n$.
 
To find the factor it is sufficient to calculate the matrixelement
$< v | {\tilde T} | v >$ for some $| v > \in R$. For this calculation
we need the properties of the Young symmetriser $Y_R$. I will give
illustrations for one of the first nontrivial examples and an
argumentation that applies to the general case.
 
Given a Young tableau one may get a vector by first labeling the boxes
{}from
one to n, which then label the  components of a tensor in $V^{\otimes
n}$. The choice here is a ``standard'' Young diagram
(\cite{Tung}p.66) (see figure \ref{figure19}).
 
 
\psbild{h}{figure19}{1.5cm}{}{example: standard Young diagram}

A vector in R can then be found by choosing an appropriate vector in
$V^{\otimes n}$. The coordinates (in terms of a basis in V) should be
different for the indices in one column, because the Young symmetriser
antisymmetrises the columns, eg. $| 1 , 2 , 3 ; 1 >$.
 
$Y_R$ then generates $| v >$ in two steps: First the columns are
antisymmetrised, i.e.:
\begin{displaymath}
a_{Y_n} = ( 1 2 3 ) ( 4 ) + ( 2 3 1 ) ( 4 ) + ( 3 1 2) ( 4 )
\end{displaymath}
\begin{displaymath}
-( 2 1 3 ) ( 4 ) - ( 1 3 2 ) ( 4 ) - ( 3 2 1 ) ( 4 )
\end{displaymath}
\begin{displaymath}
a_{Y_n} | 1 , 2 , 3 ; 1 > = | 1 , 2 , 3 ; 1 > + | 2 , 3 , 1  ; 1 > + |
3 , 1  , 2 ; 1 >
\end{displaymath}
\begin{displaymath}
- | 2 , 1 , 3 ; 1 > - | 3 , 2 , 1 ; 1 > - | 1 , 3 , 2 ; 1 >
\end{displaymath}
Transpositions are labeled by the ``boxes they transpose''. The result
of the antisymmetrisation is a vector which is antisymmetric under
transpositions within a column. This antisymmetry is indicated by
links between the antisymmetrised boxes (see figure\ref{figure20}).


\psbild{h}{figure20}{1.5cm}{}{antisymmetrised vector}
 
\newpage

Next the rows are symmetrised:
\begin{displaymath}
s_{Y_n} =   ( 1 4 ) ( 2 ) ( 3 ) + ( 4 1 ) ( 2 ) ( 3 )
\end{displaymath}


$| v >$ is given by
\begin{displaymath}
s_{Y_n} a_{Y_n} | 1 , 2 , 3 ; 1 > = | v >
\end{displaymath}
diagrammaticly:(see figure \ref{figure22})
 

\psbild{h}{figure22}{6cm}{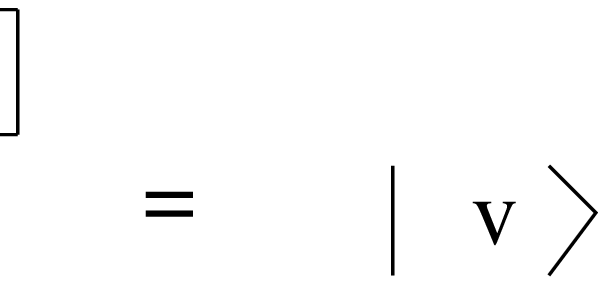}{vector in IRREP}
 
In the general case the resulting vector is a sum of $ n_1 ! ... n_l
!$ sums, each of which is symmetric under transpositions within rows
and antisymmetric under a different set of $\sum
\frac{c_i ( c_i - 1 )}{2}$ transpositions, as indicated in the
diagram.
 
In the following  $| v >$ is assumed to be normalised, i.e. $< v | v>
= 1$.
 
To evaluate $< v | {\tilde T} | v >$ we have to consider three cases:
\begin{itemize}
\item The $\sum \frac{n_i ( n_i - 1 ) }{2}$ transpositions within rows
 produce each a number $+1$ by the rearrangement lemma since
$s_{Y_n}$ is the sum of {\em all } such transpositions.
\item Each term in  $| v >$ has $\sum
\frac{c_i ( c_i - 1 )}{2}$ ``vertical'' transpositions it is
antisymmetric under. Summing over all vertical transpositions and
gathering only the terms with a definite antisymmetric behaviour leads
to $ - \sum
\frac{c_i ( c_i - 1 )}{2} | v >$ such that the contribution from this
side is:
\begin{displaymath}
 - \sum \frac{c_i ( c_i - 1 )}{2}
\end{displaymath}
\item The action of vertical transpositions on terms with no definite
symmetry makes the following contribution:
 
Because there is no definite symmetry/ antisymmetry the transposition
follows neither a row nor a link. Necessarily at least one of the
links
ends up being ``horizontal'', i.e. lying in a row (see figure \ref{figure24}). Using this
antisymmetry the contribution to the matrix element is zero (see
figure \ref{figure25}).
\end{itemize}



\psbild{h}{figure24}{4cm}{}{``horizontal links''}


\psbild{h}{figure25}{6cm}{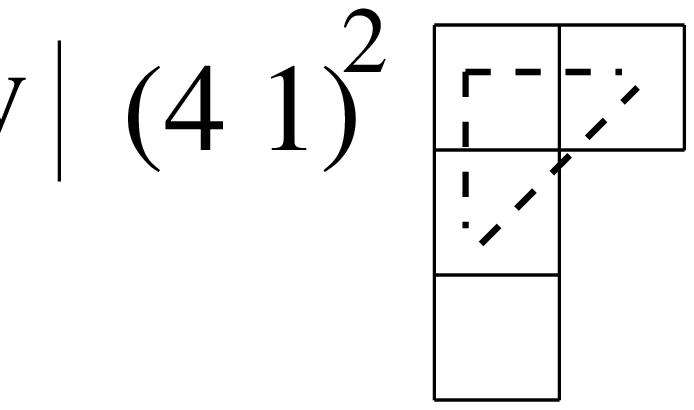}{vanishing contributions}
 
Thereby we have found:
\begin{displaymath}
< v | {\tilde T} | v > = \sum_i \frac{n_i ( n_i - 1 ) }{2} \ - \sum_i
\frac{c_i ( c_i - 1 )}{2}
\end{displaymath}
\begin{equation}
= \frac{1}{2} \sum_i  n_i^2  -  c_i^2
\end{equation}
 
This leads to
\begin{equation}
tr( {\tilde T} ) = \frac{n ( n - 1 )}{2} \chi_R (T) = \frac{{\tilde C}
(R) d_R}{2}
\end{equation}
 
This proves: $i ! \zeta_{g G}^{n i}$ is given by our symmetry factor.
For the sum over coverings the symmetry factor will be written as:
\begin{displaymath}
\sum_{\nu \in \Sigma ( G , n , i )} \frac{1}{| S_\nu |}
\end{displaymath}
 
\newpage

\section{``Tubes'' and ``contracted handles''}
 
What is still possible to do within this chapter is to interpret the
contribution from the last term in $C_2 (R)$: $- \frac{n^2}{N}$.
The ideas for this come from J.Minahan \cite{JM}.
 
We rewrite the argument of the exponential as:
\begin{displaymath}
\frac{n \lambda A}{2 N^2} + \frac{n (n - 1)}{2}  \frac{\lambda A}{N^2}
\end{displaymath}
 
The first term is interpreted as an ``infinitesimal
handle''(\cite{JM}) or ``collapsed'' handle attached to the covering
space and mapped entirely on a single point in the target space. This
object
increases the genus of the covering space by one (referring to the
square of N), gives a factor of $n \lambda A$ for the choice of its
position and a half for the indistinguishability of the two ends of
the handle.
 
The second term is viewed as a ``tube'' connecting two sheets of the
covering space: This increases the genus by one (due to the ``hole''
created by the tube) and for the choice of its position we have
$\frac{n (n - 1)}{2}  \lambda A$ (a tube relates to a transposition in
this sense).
 
These tubes are ``equivalent'' to the coincidence of two decomposition
points. They are orientation preserving, i.e. they do not
interfer with the choice of some orientation for the covering space.
Increasing the number of tubes/ handles and counting in the right
combinatorical way leads to the exponentiation of the term. (Tubes and
handles are ``local''.)

\chapter{Interpretation:  Nonchiral sum}
 
So far there has been no reason for using the orientability of the
covers in an explicit way: All terms could be interpreted consistently
without mentioning any orientation.
 
The ``chiral'' theory is the exact solution of the partition function;
but as the following examination will show it is not what we are going
for:
in the $N \rightarrow \infty $ limit the series does not give the
leading contributions first. A more useful series will be developed by
an approximate rearrangement of the chiral sum yielding the
``nonchiral'' sum. The procedure introduces a double sum over
representations (two chiral ``sectors'') instead of the single one we had
before and an ``additional'' term in the exponential weight which
couples the two sectors, which are else (approximately) independent.
The interpretation of this  coupling term ( the ``orientation
reversing tube'') makes  use of
two ``different'' orientations on the two covers each belonging to one
chiral sector.
 
\section{Leading ``composite'' representations}
 
In the $N \rightarrow \infty $  limit the weight in the sum is the
exponential: if the argument in the exponential is of order N the
contribution is damped as $exp( - O(N))$ \cite{Cor1}. While if the
argument is of order 1/N the main contribution comes from the zeroth
order
term. So it seems the most ``interesting'' contributions to the theory
come from representations which produce arguments of order $N^0$.
 
So it might be instructive to examine the weight once again:
\begin{displaymath}
exp( -\frac{\lambda A}{2 N} C_2 (R)) = exp( -\frac{\lambda A n}{2}) \
exp( -\frac{\lambda A}{2 N} {\tilde C} (R))
\ exp( \frac{\lambda A n^2}{N^2})
\end{displaymath}
The first and the third term tell that $n \approx O(N)$ is most
interesting. The second term is the one which establishes the key of
the interpretation, the connection to ``branched coverings''. To make
this one interesting we have to have ${\tilde C} (R) \approx O(N)$.
 
${\tilde C} (R)$ is essentially given by the lengths of the columns
compared with the length of rows. The columns are bounded by $N - 1$.
In a constellation where the columns have length of  $O(N)$ and where
we do not have
too many columns in the diagram we will have good contribution from
all terms. Those are easily found as ``composite representations''.
 
\vspace{1em}
 
We take two ``small'' representations R, S. ``Small'' refers to the
comparison of the parameters specifying the representation with the
gauge group parameter N.
 
We build the conjugate representation ${\bar S}$ of $S$ and combine
the rows of R and ${\bar S}$. The result is the composite
representation
$T = {\bar S} R$ (see figure \ref{figure26}).

\newpage
 
$S$ may have columns of length $\{ {\tilde c}_i \} , \sum {\tilde c}_i
={\tilde n} $, R of $\{ { c}_i \} , \sum { c}_i
={ n} $ then the columns of T are given as
\begin{displaymath}
\kappa_i = N -  {\tilde c}_{{\tilde n}_1 + 1 - i} \quad \mbox{for} \ i
\leq {\tilde n}_1
\end{displaymath}
\begin{equation}
= c_{i - {\tilde n}_1} \quad \mbox{for} \ i >  {\tilde n}_1
\end{equation}
 
%
 
\psbild{h}{figure26}{4cm}{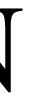}{composite representation}

We observe that all ``interesting'' IRREPs will look that way and that
putting small representations $R$, $S$ at ``different ends'' of $T$
means that their variation produces independently new composite
representations. Because we are looking for a perturbation theory we
may sum over $R$, $S$  independently as long as the leading
contributions come soon, i.e. $ N = 3$ is large enough.
 
What makes the composite representations really interesting is the
fact that the weight in the sum almost (approximately) factorises,
only giving one more, quite simple coupling term:
\begin{itemize}
\item Using the column-form of the Casimir-eigenvalue we get by simple
substitution:
\begin{equation}
C_2 (T) = C_2 (R) + C_2 (S) + \frac{2 {\tilde n} n}{N}
\end{equation}
 

\newpage

\item The dimension is not that easy to determine, but using the
hook-formula (eg. \cite{GTM129}p.78) and observing that $dim( {\bar S}
) = dim( S )$ by similarity it is easy to show that:
\begin{equation}
dim( T ) = dim( R )dim( S ) Q( O ( N^0 ) )
\end{equation}
where $Q( O ( N^0 ) )$ is a correction term which has the limit $1$
for $N \rightarrow \infty $. The zeroth order is all what will be used
in the subsequent treatment\footnote{ D.Gross and W.Taylor give the formula
\begin{displaymath}
Q = \prod_{i , j} \frac{( N + 1 - i - j ) ( N + 1 - i - j + n_i +
{\tilde n}_j )}{( N + 1 - i - j + n_i ) ( N + 1 - i - j + {\tilde n}_j
)}
\end{displaymath}
\begin{displaymath}
\approx 1 - \frac{n {\tilde n}}{N^2} + \frac{n {\tilde C} (S) +
{\tilde n} {\tilde C} (R)}{N^3} + O ( N^4 )
\end{displaymath}
where the product is ver the rows of $R$, $S$, but make only use of
the zeroth order. Furthermore in \cite{GT2} they announce a ``more
useful''
formula for the dimension.}.
\end{itemize}
 
Following this dicussion we write the partition function as:
\begin{displaymath}
Z ( G , \lambda A , N ) \approx \sum^{'}_n \ \sum^{'}_{\tilde n}
\ \sum^{'
}_{R \in Y_n} \ \sum^{'}_{S \in Y_{\tilde n}}
\end{displaymath}
\begin{displaymath}
dim( S ) \ exp(- \frac{ \lambda A }{2 N} C_2 (S)) \quad dim( R ) \
exp(- \frac{ \lambda A }{2 N} C_2 (R))
\end{displaymath}
\begin{equation}
exp(- \frac{ \lambda A n {\tilde n}}{ N^2 } )
\end{equation}
where the prime indicates that the approximation will only be
good for summations over small representations. In case N is not large
(i.e. ``finite'') this summation is a massive overcounting (cf.
\cite{Baez})

\section{Orientation reversing tubes}
 
Each chiral sector is already interpreted. All that has to be done
here is to provide an interpretation for the ``coupling term''
 
The term, of course, looks like the one we interpreted as a tube in
the chiral sum. But taking a tube like in the chiral theory is not
consistent with the idea of composite representations given by two
independent sectors. This forbids the usage of perturbative tools from
the sectors.
 
In the topology of surfaces the ``connection'' of two surfaces is
usually done  by removing a disc from both and connecting the
resulting one-spheres with a cylinder. This procedure has the
advantage that the result, the ``connected sum'', is again a surface
and has the same orientability character as the cartesian product
(\cite{Naka}p.58). But we do not want to do algabraic topology, we
want to find a suitable, consistent formalism.
 
What the cylinder in the connected sum actually does is to map one
one-sphere onto the other in a continuous way. This looks like a
``homotopy''. As long as our ``homotopy'' is locally diffeomorphic
everywhere we will not be able to give an interpretation different
{}from
the tubes used for the chiral sectors.
 
\newpage

Taking the cylinder $C = S^1 \times [ 0 , 1 ] = \{ (z , x): |z| = 1 ,
0 \leq x \leq 1  \}$
and comparing the maps
\begin{displaymath}
\nu_- (z , x) = z ( 1 - 2 x)
\end{displaymath}
\begin{displaymath}
\nu_+ (z , x) = z ( 1 - 2 x) \quad \mbox{for} \ x \leq \frac{1}{2}
\end{displaymath}
\begin{displaymath}
\qquad = {\bar z} (2 x -1) \quad \mbox{for} \ x \geq \frac{1}{2}
\end{displaymath}
we can illustrate what a orientation preserving and a orientation
reversing tube is:
\begin{itemize}
\item$\nu_+$:
\newline z indicates some sense of circulation in the complex plane,
${\bar z} $ the opposite one . In the middle ($x = \frac{1}{2}$) we
done not have any sense of circulation; the orientation switches its
sign.
\item $\nu_-$:
\newline Here the orientation remains the same all the way.
\end{itemize}
Geometrically (see figure \ref{figure27}):

 
\psbild{h}{figure27}{5cm}{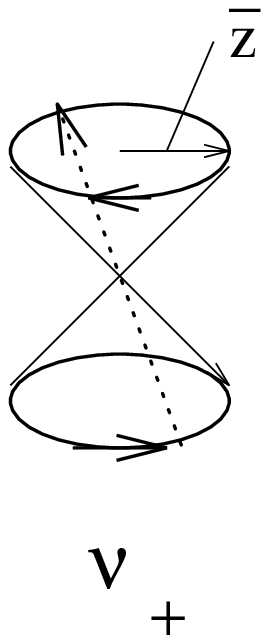}{orientation reversing tubes}

As before one imagines the genus to be increased by one and the
combinatorical factor is, of course, $n {\tilde n} \lambda A$. Nice is
that each such tube carries a minussign. Because these orientation
reversing tubes are local they exponentiate.
 
\newpage

\section{Free energy}

Be $\Sigma_G$ the set of disconnected covering maps then we have:
\begin{displaymath}
Z ( G , \lambda A, N ) =
\end{displaymath}
\begin{displaymath}
 \sum_{\nu \in  \Sigma_G} \frac{(-)^{\tilde
t}}{| S_\nu |} \ \ exp( - \frac{ n \lambda A}{2} )
 \ \ \frac{(\lambda A)^{( i + t + {\tilde
t} + h)}}{i !  \ t ! \  {\tilde t} ! \  h !} \quad \times
\end{displaymath}
\begin{equation}
N^{ n ( 2 - 2 G ) - i - 2 ( t + {\tilde t} + h)} \ \ [ 1 + O (
\frac{1}{N}) ]
\end{equation}
where n is the winding number, i: ``branching'' number, t: number of
orientation preserving tubes, ${\tilde t}$: number of orientation
reversing tubes, h: number of infinitesimal handles, $\frac{1}{| S_\nu
|}$: the symmetry factor, $ O (\frac{1}{N}) $: subleading
contributions from the dimension (only for $G \neq 1$).
 
Some argumentation like in QFT may apply to show that the free energy
is the corresponding sum of connected coverings (${\tilde \Sigma}_G$)
only:
\begin{displaymath}
ln Z = W ( G , \lambda A, N ) =
\end{displaymath}
\begin{displaymath}
\sum_{\nu \in  {\tilde \Sigma}_G} \frac{(-)^{\tilde
t}}{| S_\nu |} \ \ exp( - \frac{ n \lambda A}{2} )
 \ \ \frac{(\lambda A)^{( i + t + {\tilde
t} + h)}}{i !  \ t ! \  {\tilde t} ! \  h !} \quad \times
\end{displaymath}
\begin{equation}
N^{ n ( 2 - 2 G ) - i - 2 ( t + {\tilde t} + h)} \ \ [ 1 + O (
\frac{1}{N}) ]
\end{equation}

\chapter{Frobenius' formulae,  boundaries \& gluing}
 
The content of this chapter is already a little bit beyond the scope
of this work, because it rather the subject of D.Gross' and W.Taylor's
second article \cite{GT2} than of their first one. Nevertheless I
shall outline the development at the end of their first article.
 
The concepts used in this chapter have already been introduced and
examplified before:
\begin{itemize}
\item The ``holonomies'' associated with boundaries.
\item The ``gluing'' of partition functions of areas with boundaries
along edges using the orthogonality of the characters.
\item The way of interpreting the 1/N expansion using branched
coverings, calculating the symmetry factor, attaching tubes and
handles.
\end{itemize}
 
For the discussion of coverings of bordered surfaces the introduction
of a new set of class functions (the ``power sums'' \cite{GTM129}, the
``loop variables'' \cite{Cor1}) is convenient. This will be the
subject of the first section.

\section{Frobenius' formulae}
 
In the further development of the stringy approach to $QCD_2$ another
set of class functions is used (see eg. \cite{Baez} \cite{GT2}
\cite{Cor1} \cite{Cor2}):
 
Be $\sigma \in S_n$ with cycles of length $\{ n_i \}_{i \in k}$, $\sum
n_i = n$ then it is defined:
\begin{equation}
\Upsilon_\sigma (U) \equiv \prod_{j = 1 }^k (tr U^j )^{n_j}
\end{equation}
These functions, in \cite{Cor1} referred to as ``loop variables'',
form
a complete set of class functions (see ``gluing'' formula later on).
 
They are a form of the ``power sums'' defined in eg.\cite{GTM129}p.48;
because $U \in U (N)$ is diagonisable and the trace is invariant under
similarity transformations they are, in effect, the same.
 
The ``loop variables'' $\Upsilon_\sigma$ satisfy relations with the
characters $\chi_R$, called the ``Frobenius relations'':
\begin{equation}
\Upsilon_\sigma (U) = \sum_R \chi_R (\sigma )  \chi_R (U)
\end{equation}
This is a form of the Frobenius character formula
\cite{GTM129}p.49\footnote{ The connection between the ``Schur
polynomials'' and the characters $ \chi_R (U)$ is given in theorem
6.3, same reference}.
 
There is an ``inverse'' relation, too:
\begin{equation}
\chi_R (U) = \sum_{\sigma \in S_n} \frac{1}{n !} \chi_R (\sigma )
\Upsilon_\sigma (U)
\end{equation}
which can be found by \cite{GTM129}p.460\& 534.
 
Using these relations, some identities stated in the proof of the
connection
${\tilde C} (R) \leftrightarrow \chi_R ( T  )$ we readily find the
``gluing'' formulae for loop variables as eg.
\begin{displaymath}
\int dU \ \Upsilon_\sigma (U) \ \Upsilon_\tau (U^+ )
= \sum_R  \chi_R (\sigma ) \ \sum_{R^{'}}  \chi_{R^{'}} (\tau ) \int
dU \
\chi_R (U) \ \chi_{R^{'}} ( U^+ )
\end{displaymath}
\begin{equation}
= \sum_R  \chi_R (\sigma ) \chi_R (\tau ) = \ \delta_{[ \sigma ] , [
\tau ]}  \ \frac{n !}{| [ \sigma ] |}
\end{equation}
where $[ \sigma ]$ is the conjugacy class of $ \sigma $ and $| [
\sigma ] |$ is the number of elements in $[ \sigma ] $.
 
For the treatment of the nonchiral sum the loop variables are
generalised for composite representations (see \cite{Cor1}5.5
\cite{GT2}App.1). A thorough discussion of the interpretation of the
nonchiral sum in terms of ``coupled loop functions'' is not possible
in this work.
 
But in the the interpretation of the chiral sum the loop variables
prove to be particularly well suited to the discussion of coverings
with boundaries. This is basicly because ``during'' the covering map
the boundaries of the covering space wind around the boundaries of the
target space several times, which is reflected by the different powers
in the definition of the loop variables $\Upsilon_\sigma (U)$.
 
It is, therefore, necessary to express the partition function in terms
of the loop variables and then to give a matching interpretation.

\section{Interpretation:  branched covering of a plaquette}
 
The partition function of a plaquette with area A and holonomy U is:
\begin{equation}
Z_\Delta ( U ) = \sum_n \sum_{R \in Y_n}^{'} \ dim(R) \ \chi_R (U) \
exp(-
\frac{\lambda A}{2 N} C_2 (R))
\end{equation}
The following treatment is very analogous to the interpretation of the
partition function for closed surfaces with genus G. Because the
term$-\frac{n^2}{N}$ in the Casimir- eigenvalue does not give rise to
any new argumentation it will be dropped for this discussion.
 
\newpage

So the 1/N- expansion we have to interpret is:
\begin{displaymath}
Z_\Delta ( U ) = \sum_n  \ exp(-
\frac{n \lambda A }{2 }) \ \sum_i \frac{(\lambda A)^i }{i !} \quad \times
\end{displaymath}
\begin{equation}
\sum_R [ \frac{d_R}{n !} \ (-\frac{{\tilde C} (R)}{2} )^i \ \chi_R (U)
 \ N^{n - i} + O( N^{n - i - 1})]
\end{equation}
Inserting the Frobenius character formula:
\begin{displaymath}
Z_\Delta ( U ) = \sum_n  \ exp(-
\frac{n \lambda A }{2 }) \ \sum_i \frac{(\lambda A)^i }{i !} \quad \times
\end{displaymath}
\begin{equation}
\sum_\sigma [ \frac{(-)^i}{| [ \sigma ] |} \ ( \sum_R \frac{d_R | [
\sigma ] |}{n !^2} (\frac{{\tilde C} (R)}{2} )^i \  \chi_R (\sigma )
\Upsilon_\sigma (U) \ N^{n - i} ) + O( N^{n - i - 1}) ]
\end{equation}
We define:
\begin{equation}
\phi _{n i }^\sigma \equiv \sum_R  \frac{d_R | [
\sigma ] |}{n !^2} (\frac{{\tilde C} (R)}{2} )^i \  \chi_R (\sigma )
\end{equation}
As for areas with no boundaries this quantity will be interpreted as
sum over branched covers with the number of distinct honomorphisms as
symmetry factor.
 
\vspace{1em}
 
The classification of surfaces with boundaries was described in
chapter 6. With this the discussion for branched coverings of $\Delta$
is as before:
We define $\Sigma_\sigma (n , i )$ to be the set of n-fold covers of
$\Delta$ with ``branching'' number i and and boundary of k disjoint
one spheres $S^1$ which cover the boundary of $\Delta$ $\{ n_i \}$
times. The $\{ n_i \}$ are the sizes of the $k$ cycles in the permutation
$\sigma$.
 
Just as in chapter 6:
\begin{displaymath}
\sum_{\nu \in \Sigma_\sigma (n , i )} \frac{1}{| S_\nu |} =
\sum_{\tau_1 .. \tau_i \in T} \ \frac{| [ \sigma ] |}{n !}  \delta(
\tau_1 .. \tau_i \sigma )
\end{displaymath}
\begin{equation}
= \sum_R \frac{d_R | [ \sigma ] |}{n !^2} (\frac{{\tilde C} (R)}{2}
)^i  \  \chi_R (\sigma )
\end{equation}
 
We can now rewrite the partition function as sum over coverings:
\begin{equation}
Z_\Delta ( U ) = \sum_\nu [ \frac{(-)^i}{| S_\nu |} \ exp(-
\frac{n \lambda A }{2 }) \ \frac{(\lambda A)^i }{i !} \prod_j (tr
{\hat
U}_j) \ N^{n -i} + O( N^{n - i - 1} ) ]
\end{equation}
where the ${\hat U}_j$ are the pullback-holonomies of the gauge field
to the covering space \footnote{for pullback of holonomies see eg.
\cite{Naka}p.45\& 201}. The ${\hat U}_j$ are the reason for using the
loop variables rather than the characters in the partition function.

\section{Some outlooking remarks:  On gluing in covering
space}

The discussion of the previous section may serve as an example how the
covering of bordered
surfaces works. By gluing plaquettes we get more complicated surfaces.
In order to extent this discussion to the general case one has to
determine the behaviour of coverings under gluing of the target space.
In any case it is necessary to find a formalism for the gluing  in
order to
make the interpretation complete because the invariance under
subtriangulation/ the additivity of the partition function is one of
the essential features of $QCD_2$. The solution to this problem is the
subject of the extensive work published in \cite{GT2} and \cite{Cor2},
and will therefore not be given here. Nevertheless I  want
to outline where about the next step leads.
 
Gluing involves integrals like
\begin{displaymath}
\int dU \ \frac{| [ \sigma ] |}{n !} \ \Upsilon_\sigma (V U) \ \frac{|
[ \tau ] |}{n !} \ \Upsilon_\tau (U^+ W)
\end{displaymath}
\begin{displaymath}
= \sum_{\sigma^{'} \in  \ [ \sigma ] } \ \sum_{\tau^{'} \in \  [ \tau
] } \
\sum_R \ \frac{\chi_R ( \sigma^{'} ) \chi_R ( \tau^{'} )}{dim(R)}
\chi_R (V W)
\end{displaymath}
\begin{equation}
= \sum_{\sigma^{'} \in  \ [ \sigma ] } \ \sum_{\tau^{'} \in  \ [ \tau
] }
\ \sum_{\alpha \in S_n} \ N^{- n } \ \delta ( \sigma^{'} \tau^{'}
\alpha \Omega_n^{- 1}) \ \frac{\Upsilon_\alpha ( V W )}{n !}
\end{equation}
where
\begin{displaymath}
\Omega_n = \sum_{\mu \in S_n} (\frac{1}{N})^{n - K_\mu} \ \mu
\end{displaymath}
and $K_\mu$ is the number of cycles in $\mu$
(\cite{Cor1}(5.13)(5.34)).
 
$\Omega_n $ has to be introduced to express the partition function
once again in terms of loop variables.

The gluing of edges of the cover is associated with the closely
related integral \cite{Cor1}(5.36)\footnote{The formula for the  proof
of the character
gluing formula in chapter 2 is this one for $n = 1$}
\begin{equation}
\int dU \ U_{j_1}^{i_1} .. U_{j_n}^{i_n}{U^+}_{l_1}^{k_1} ..
{U^+}_{l_n}^{k_n} = \sum_{\rho \sigma} \frac{\delta (\Omega_n^{- 1}
\rho )}{N^n} \ \delta_{l_{\sigma ( 1 )}}^{i_1} \delta^{k_{(\rho
\sigma)
( 1 )}}_{j_1} .. \delta_{l_{\sigma ( n )}}^{i_n} \delta^{k_{(\rho
\sigma)
( n )}}_{j_n}
\end{equation}

The $\delta_\beta^\alpha $ express the gluing of an edge of the sheet
$\alpha$ with one of the sheet$\beta$.
 
The $\Omega_n^{- 1} $ are then objects of new interpretation
procedures (\cite{Cor1}\cite{Cor2}\cite{GT2}).
 
\vspace{1em}
 
The nonchiral sum includes a new ``interaction'' between the two
sectors due to the fact that $\chi_{{\bar S} R} (U) \neq \chi_S (U^+)
\chi_R (U)$.
 
Because the conjugate representation ${\bar S}$ is associated with
traces over the adjoint of the holonomies $U^+$ the loop variables
provide a further understanding how the two different orientations
come into the theory.
 
D.Gross and W.Taylor \cite{GT1} argue that the interaction term $\sim
\frac{n {\tilde n}}{N}$ would cancel the contributions from folds
occurring during the gluing edges of sheets with opposite orientation.
These matters are furtherly discussed in the later publications.

\newpage

\section{Outview}
 
It proved to be impossible to give a ``complete'' account of the
progress made in this direction so far. Still I want to indicate where
else results may be found.
 
The program was continued by D.Gross and W.Taylor in \cite{GT2} and
S.Cordes et al. in \cite{Cor2}. The latter seem to have been able to
connect the interpretation with topological string theory. Furthermore
they announce another approach to the string theory interpretation
using harmonic covering maps rather than holomorphic ones by P. Horava
\cite{Horav}.
 
Interesting papers on $YM_2$ are for example one by Blau \& Thompson
\cite{Blau} and  one by Witten \cite{Witt}. In the latter $YM_2$ is
discussed on nonorientable surfaces.
 
In the 1980's much effort has been made to derive a resonable
1/N-expansion in equations for loops (``loop dynamics'') See for
example \cite{Mig1}.
 
The high energy behaviour of strings is discussed eg. by D.Gross and
P.Mende \cite{GM}. They find a gaussian falloff at high energies
which is incompatible with the observed powerlike behaviour of strong
cross sections at high energies.
 
Nevertheless Verlinde \& Verlinde made a recently updated effort to
describe high energy QCD in terms of two dimensional nonabelian shock
waves
\cite{Verl}.
 
Finally I would like to mention the work by J.Baez and W.Taylor
\cite{Baez}. They give a description of $QCD_2$ for finite N by usage
of loop variables and a secondly quantised Hamiltonian formalism.

\chapter*{}
\section*{Conclusions}

Free $QCD_2$ on a compact surface is solved exactly. The 1/N-expansion
of the  related
partition function for closed surfaces can be given as a ``string''
theory.
This formalism can incorporate boundaries.

The question whether 1/N is a good perturbative parameter can not be
answered
before a generalisation to higher dimension and explicit calculations
of eg. bound state energies.

\section*{Acknowledgement}

I am indebted to P.Mansfield  for his helpful
supervision, and the Centre for Particle Theory for the cooperative
climate and the facilities I could use.

I am grateful to the ``Studienstiftung des Deutschen Volkes'' for their
financial
and administrative support and to my family who made my being here
possible.

I would like to thank J.Brodzki (UNI Durham), and G.Br\"u{}chert (UNI
Hamburg) for helpful
discussions on group theory, U.Harder (UNI Durham), 
and J.Gladikowski (UNI Durham) for their advise while managing
\LaTeX{}, and U. Harder (UNI Durham), W.Oxbury (UNI Durham), 
A.Taormina (UNI Durham), K.Kunze
(UNI Sussex), and A.B\"a{}cker (UNI Hamburg) for making some of the
references accessible to me.

\end{document}